\documentclass[reprint,groupedaddress,aps,pra,groupedaddress,amsmath,amssymb,showpacs]{revtex4-1}
\usepackage{booktabs,ulem}
\usepackage{graphicx,amsmath,amssymb,epsfig,color,epstopdf,comment,float}
\usepackage{color}
\usepackage{dcolumn}
\usepackage{cleveref}
\usepackage{bm}
\allowdisplaybreaks
\begin{document}

\title{Switch and Phase Shift of Photon Polarization Qubits via Double Rydberg Electromagnetically Induced Transparency}
\thanks{The article has been accepted and published by Physical Review A.}%

\author{Yao Ou$^{1}$}
\author{Guoxiang Huang$^{1,2,3,}$}\thanks{gxhuang@phy.ecnu.edu.cn}
\affiliation{
$^1$State Key Laboratory of Precision Spectroscopy, East China Normal University, Shanghai 200241, China \\
$^2$NYU-ECNU Joint Institute of Physics, New York University Shanghai, Shanghai 200062, China \\
$^3$Collaborative Innovation Center of Extreme Optics, Shanxi University, Taiyuan 030006, China
}

\date{\today}

\begin{abstract}
We propose and analyze a scheme for manipulating the propagation of single photon pulses with two polarization components in a Rydberg atomic gas via double electromagnetically induced transparency. We show that by storing a gate photon in a Rydberg state a deep and tunable potential for a photon polarization qubit can be achieved based on strong Rydberg interaction. We also show that the scheme can be used to realize all-optical switch in dissipation regime and generate a large phase shift in dispersion regime for the photon polarization qubit. Moreover, we demonstrate that such a scheme can be utilized to detect weak magnetic fields. The results reported here are not only beneficial for understanding the quantum optical property of Rydberg atomic gases, but also promising for designing novel devices for quantum information processing.
\end{abstract}

\maketitle

\section{Introduction}\label{Sec1}

Photons do not interact with each other in vacuum, and also hardly interact with their environments. The linear (or nearly linear) property of light propagation, in combination with high speed, large bandwidth, and low loss, has made photons be excellent information carriers for optical communications over long distances. However, for quantum information processing strong interactions between photons are required. Although interactions between photons may be obtained through some nonlinear optical processes~\cite{Boyd2008}, optical nonlinearities
realized through these processes are too weak for all-optical quantum information processing.

All-optical switch is a photonic device by which a gate pulse can effectively change the transmission of a target pulse without the aid of electronic techniques. For quantum information processing, it is desirable to build single-photon switches in which the gate pulse contains only one photon. However, building single-photon switches is generally difficult, due to the reason that Kerr nonlinearities in conventional optical media are too small at single-photon levels. Nevertheless, the research of electromagnetically induced transparency (EIT)~\cite{Fleischhauer2005} in past three decades has triggered the possibility for realizing strong optical nonlinearities at few-photon levels~\cite{Chang2014}.

Among a wide variety of physical systems that support EITs,
Rydberg atomic gases~\cite{Gallagher2008,Saffman2010} are particularly attractive, in which strong atom-atom interaction can be effectively mapped onto strong  photon-photon interaction via Rydberg-EIT~\cite{Friedler2005,Mohapatra2007PRL,Adams2010}.
In recent few years, tremendous attention has been paid to the study on various single- and few-photon states and their quantum dynamics in atomic gases working under the condition of  Rydberg-EIT~\cite{Gorshkov2011PRL,Firstenberg2013,He2014,Bienias2014,Caneva2015,
Maghrebi2015,Li2015,Gullans2016,Murray2016,Thompson2017, Pohl2019,Jachymski2016,Das2016,Yang2016,Gullans2017,Moos2017,Liang2018,
Cantu2020,Bienias2020,Ou2022,Drori2023,DingY2023,Murray2016adv,Adams2020}. Especially, many single-photon devices (including single-photon switches and phase gates), which are promising for all-optical quantum information processing,
have been demonstrated experimentally~\cite{Dudin2012,Peyronel2012,Baur2014PRL,Gorniaczyk2014,Tiarks2014PRL,
Gorniaczyk2016,Tiarks2016,Ripka2018,Tiarks2019,Ornelas-Huerta2020,Vaneecloo2022,
Stolz2022,Shi2022,Ye2023,Murray2016adv,Adams2020}.

In this article, we suggest a scheme to realize a new type of single-photon switch based on strong Rydberg interaction. Different from those explored before, in which single-photon switches were designed for photon states with only one polarization component $|\sigma\rangle$~\cite{Li2015,Murray2016,Pohl2019,Baur2014PRL,Gorniaczyk2014,Tiarks2014PRL,
Gorniaczyk2016}, in our scheme the single-photon switch is for the photon state with two polarization components ($\sigma^+$ and $\sigma^-$), i.e. for photon polarization qubit $c_+ |\sigma^+\rangle+c_- |\sigma^-\rangle$ ($c_+$ and $c_+$ are complex constants satisfying $|c_+|^2+|c_-|^2=1$). The system we consider is a cold Rydberg atomic gas working under the condition of double Rydberg-EIT. Recently, such an EIT has been used  to acquire large self- and cross-Kerr nonlinearities and some novel nonlinear optical phenomena (e.g., giant magneto-optical rotation, self-organized optical spatial structures, and Stern–Gerlach deflection of light bullets) for situations with large probe photon
numbers~\cite{Sinclair2019,Mu2021,Shi2021PRA,Shi2021OL,Mu2022}. In contrast with
these works, where semi-classical approaches were used, in the present study the probe-laser field in the system is assumed to be in a single-photon state, and hence an all-quantum approach for both the atoms and the probe field is needed.

We shall show that, by storing a gate photon in a Rydberg state, a deep and tunable optical potential (called Rydberg defect potential below) for a photon polarization qubit can be prepared through the strong Rydberg interaction. We also show that by using this scheme it is possible to design effective switch for the photon polarization qubit if the system works in dissipation regime. Moreover, large phase shifts for the two polarization components of the photon polarization qubit can be generated when the system works in dispersion regime. In addition, we demonstrate that such a scheme can be utilized to detect weak magnetic fields.
The research results reported here are useful not only for understanding the quantum optical property of Rydberg atomic gases, but also for designing novel single-photon devices promising for optical quantum information processing~\cite{Kok2010}.

The remainder of the article is arranged as follows. In Sec.~\ref{Sec2}, we describe the physical model under study and derive two-component envelope equations of the quantized probe field  based on Heisenberg-Maxwell (HM) equations. In Sec.~\ref{Sec3}, we solve the two-component envelope equations and  present analytical and numerical results on the realization of the Rydberg defect potential,  polarization qubit switch, phase shifts of the two qubit components, and magnetic-field-induced switching behavior for the polarization qubit. Finally, in Sec.~\ref{Sec4} we summarise the main results obtained in this work.

\section{Model and equations of motion}\label{Sec2}
\subsection{Physical model}\label{Sec21}
\begin{figure*}
\centering
\centerline{\includegraphics[width=2.1\columnwidth]{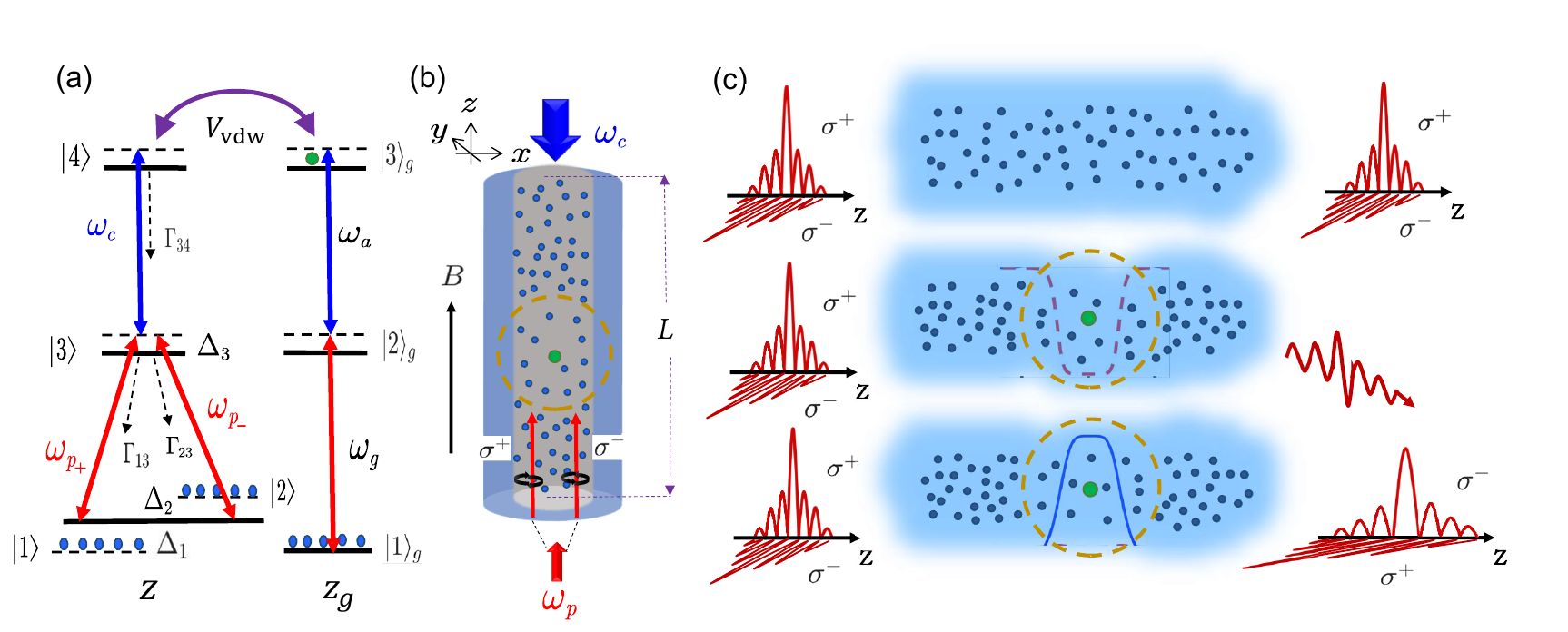}}
\caption{\footnotesize Schematics of the model and the propagation of the single photon polarization qubit.
(a)~Left part: level diagram and excitation scheme of the double Rydberg-EIT,
consisting of two ladder-shaped EIT excitation paths, i.e. $|1\rangle \leftrightarrow |3\rangle \leftrightarrow |4\rangle$ and $|2\rangle \leftrightarrow |3\rangle \leftrightarrow |4\rangle$, with $|1\rangle$ and $|2\rangle$ the two lower states, $|3\rangle$ the excited state, and $|4\rangle$ the Rydberg state.
Right part: a gate photon is stored in another Rydberg state $|3_g\rangle$ of the gate atom via another Rydberg-EIT through the excitation path $|1_g\rangle\rightarrow |2_g\rangle \rightarrow |3_g\rangle$. Red (blue) lines with double-headed arrows represent the probe (control) fields. The purple line with double arrows and symbol $V_{\rm vdW}$ represents the van der Waals interaction between the Rydberg atom located at the position $z$ and the Rydberg atom at the position $z_g$ (gate atom).
For detailed description of the probe and control fields, atomic decay rates $\Gamma_{\alpha\beta}$, and detunings $\Delta_{\alpha}$, see the text.
(b)~Suggested geometry of the system. The gate atom (assumed to locate at the middle of the atomic gas, i.e., $z_g=L/2$)
excited to the Rydberg state $|3\rangle_{g}$ by the gate photon is denoted by the solid green circle; other atoms are denoted by solid blue circles.
The domain centred at the gate atom forms a Rydberg blockade sphere with radius $r_b$ (indicated by the dashed yellow circle), which contributes a Rydberg defect potential for the incident probe photon qubit. An external magnetic field $B$ applied along the $z$ direction results in the detunings $\Delta_1=-\Delta_2=-\mu_{B} B/(3 \hbar)$.
(c)~Top: schematics of the free propagation of the two polarization components ($\sigma_{+}$ and $\sigma_{-}$) of the photon qubit in the absence of the stored gate photon (i.e. the qubit switch is off).
Middle: the stored gate photon induces a dissipation-type Rydberg defect potential (with a large imaginary part) indicated by the dashed purple curve, blocking the transmission of the photon qubit (i.e. the qubit switch is on).
Bottom: the stored gate photon induces a dispersion-type Rydberg defect potential (with a large real part) indicated by the solid blue curve.
Significant phase shifts are generated for the two polarization components of the photon qubit.
}
\label{Fig1}
\end{figure*}

We start to consider a cold four-level atomic gas (lower states $|1\rangle$ and $|2\rangle$, excited state $|3\rangle$, and Rydberg state $|4\rangle$) with an excitation scheme of inverted-Y-type configuration,
interacting with a weak, pulsed probe laser field (target pulse) of central wavenumber $k_p$ and angular frequency $\omega_{p}=k_p c$, and a strong, continuous-wave control laser field of wavenumber $k_c$ and angular frequency $\omega_{c}=k_c c$; see the left part of Fig.~\ref{Fig1}(a). To suppress first-order Doppler effect, the probe (control) field is assumed to propagate along  $z$ ($-z$) direction.

We assume that the probe field consists of two polarization components,
i.e. a right-circular ($\sigma^{+}$) and left-circular ($\sigma^{-}$) ones,
coupling to transitions  $|1\rangle\leftrightarrow|3\rangle$ and
$|2\rangle\leftrightarrow|3\rangle$) respectively; the control field
couples to the transition $|3\rangle\leftrightarrow|4\rangle$.
$\Gamma_{13}$, $\Gamma_{23}$, and $\Gamma_{34}$ are spontaneous decay rates from $|3\rangle$ to $|1\rangle$, $|3\rangle$ to $|2\rangle$, and $|4\rangle$ to $|3\rangle$, respectively.
$\Delta_{1}$ and $\Delta_{2}$ are Zeeman energy splitting of atomic ground state level, induced by an external magnetic field $B$ applied along the $z$-direction~\cite{Petrosyan2004}; $\Delta_{3}$ and $\Delta_{4}$ are one-photon and two-photon detunings, respectively.
The excitation scheme shown in the left part of Fig.~\ref{Fig1}(a) is the basic configuration of double Rydberg-EIT; it consists of two ladder-shaped EIT excitation paths, i.e. $|1\rangle \leftrightarrow |3\rangle \leftrightarrow |4\rangle$ and $|2\rangle \leftrightarrow |3\rangle \leftrightarrow |4\rangle$.

For simplicity, we assume that the system behaves as a one-dimensional one, which can be realized by taking a cigar-shaped atomic gas, or an atomic gas filled into a waveguide with small transverse sizes, so that the optical fields of the system in transverse directions are tightly confined, and hence the diffraction effect can be safely neglected. Thereby, a (1+1)-dimensional (i.e. time plus the space along the $z$-axis) model is sufficient to describe the dynamics of the system, as schematically shown in Fig.~\ref{Fig1}(b)~\cite{note00}.
The total electric field in the system reads
\begin{subequations}\label{Es0}
\begin{align}
& {\hat{\bf E}}(z,t)={\bf E}_{c}(z,t)+{\hat{\bf E}}_{p}(z,t) \\
& {\bf E}_{c}(z,t)={\bf e}_{c}{\cal E}_{c}e^{i(-k_{c}z-\omega_{c}t)}+{\rm c.c.},\\
& {\hat{\bf E}}_{p}(z,t)={\hat{\bf E}}_{p+}(z,t)+{\hat{\bf E}}_{p-}(z,t),\\
& {\hat{\bf E}}_{pj}(z,t)={\bf e}_{pj}{\cal E}_{p}{\hat E}_{pj}(z,t)e^{i(k_{p\pm}z-\omega_{p\pm}t)}+{\rm h.c.}.
\end{align}
\end{subequations}
Here $j=+,\,-$, $k_{p\pm}=k_p$, $\omega_{p\pm}=\omega_{p}$, and c.c. (h.c.) represents complex (Hermitian) conjugate;
${\bf e}_{c}$ and ${\cal E}_{c}$ are the unit polarization vector and amplitude of the control field; ${\cal E}_{p}\equiv \sqrt{\hbar\omega_{p}/(2\varepsilon_{0}V)}$ is the field amplitude of single probe photon, with
$V=LA_0$ is the optical volume of the system ($A_0$ and $L$ are the cross section area and longitudinal size of the atomic ensemble, respectively); ${\bf e}_{p+}=({\bf e}_{x}+i{\bf e}_{y})/\sqrt{2}$ and ${\hat E}_{p+}(z,t)$  [${\bf e}_{p-}=({\bf e}_{x}-i{\bf e}_{y})/\sqrt{2}$ and ${\hat E}_{p-}(z,t)$] are respectively the unit polarization vector and annihilation operator of probe photon for the $\sigma^{+}$\, ($\sigma^{-}$) polarized component. ${\hat E}_{p+}(z,t)$ and  ${\hat E}_{p-}(z,t)$
obey commutation relations
\begin{subequations}
\begin{eqnarray}
&& [{\hat E}_{pj}(z,t),{\hat E}_{pj'}(z',t)]=[{\hat E}_{pj}^{\dag}(z,t),{\hat E}_{pj'}^{\dag} (z',t)]=
0,\nonumber\\
\\
&& [{\hat E}_{pj}(z,t),{\hat E}_{pj'}^\dag(z',t)]=L\delta(z-z')\delta_{jj'},
\end{eqnarray}
\end{subequations}
with $j, j'=+,-$ and $L$ being the size of the system along the $z$ direction.
We also assume that the incident probe field is a single-photon pulse, and the quantum state of the pulse is a polarization qubit because the pulse contains two polarization components.

In order to design a switch for the photon polarization qubit, a gate photon must be prepared. Here we adopt the idea adopted in Refs.~\cite{Baur2014PRL,Gorniaczyk2014,Tiarks2014PRL,
Gorniaczyk2016}, i.e., before the incidence of the probe photon a gate photon is stored in another Rydberg state $|3_g\rangle$ of an atom (called gate atom). This can be realized by using another Rydberg-EIT through the excitation path $|1_g\rangle\rightarrow |2_g\rangle \rightarrow |3_g\rangle$. Here, the gate photon pulse (with central angular frequency $\omega_{g}$ and half Rabi frequency $\Omega_g$) couples the atomic states $|1\rangle_{g}$ and  $|2\rangle_{g}$, and a strong, assisted laser field (with central angular frequency $\omega_{a}$ and half Rabi frequency $\Omega_a$) couples the states $|2\rangle_{g}$ and $|3\rangle_{g}$, as shown in the right part of Fig.~\ref{Fig1}(a). In this way, the incident gate photon is stored in the gate atom, and hence the Rydberg state $|3\rangle_{g}$ can have the atomic population of unit probability.

Assume that the atom excited into the state $|4\rangle$ locates at position $z$. Because both $|3\rangle_{g}$ and  $|4\rangle$ are Rydberg states, there exists a strong Rydberg-Rydberg interaction between the gate atom (at position $z_g$) and the atom at $z$. Such an interaction can be described by the van-der-Waals (vdW) interaction potential of the form
\begin{eqnarray}\label{vdWP}
\hbar V_{\rm vdW}(z_g-z)=-\frac{\hbar C_6}{|z_g-z|^6},
\end{eqnarray}
if both $|4\rangle$ and $|3_g\rangle$ are taken to be S state.
Here $C_6$ is called dispersion coefficient. The Rydberg-Rydberg interaction results in atomic level shifts and hence induces an important phenomenon, called Rydberg blockade~\cite{Gallagher2008,Saffman2010,Murray2016adv,Adams2020}, by which only one atom can be excited to Rydberg states in the region of Rydberg blockade sphere of radius $r_b$~\cite{note0}.

Under electric-dipole, rotating-waving, and paraxial approximations, the effective Hamiltonian of the atomic ensemble is given by ${\hat H} ={\hat H}_{\rm AF}+{\hat H}_{\rm AG}$, with
\begin{subequations}\label{Hami}
\begin{eqnarray}
{\hat H}_{\rm AF} &=& -\hbar  \int_{-\infty}^{+\infty}dz \rho_a (z) \left[\sum_{\alpha=1}^{4}\Delta_{\alpha}
\hat{S}_{\alpha\alpha}(z,t)\right. \nonumber \\
&&+\Omega_c\hat{S}_{34}(z,t)+g_{p+}\hat{S}_{13}(z,t){\hat E}_{p+}(z,t)\nonumber \\
&& \left.+ g_{p-}\hat{S}_{23}(z,t){\hat E}_{p-}(z,t)+{\rm h.c.}\right], \label{Hami2}\\
{\hat H}_{\rm AG} &=& \int_{-\infty}^{+\infty} dz \rho_a (z)
\int_{-\infty}^{+\infty} dz^{\prime}_g  \rho_g (z^{\prime}_g) \nonumber \\
&& \times \left[\hat{S}_{33}(z^{\prime}_g,t)\hbar V_{\rm vdW}(z^{\prime}_g-z)\hat{S}_{44}(z,t) \right],
\label{Hami3}
\end{eqnarray}
\end{subequations}
where, ${\hat H}_{\rm AF}$ is the Hamiltonian describing the atom-light interaction, and ${\hat H}_{\rm AG}$ is the one describing the Rydberg-Rydberg interaction between atoms in the Rydberg state $|4\rangle$ and gate atoms in the Rydberg state $|3\rangle_{g}$. In these expressions, $\rho_{g}$ is the linear density of gate atoms; $\rho_{a}$ is the linear density of atoms other than the gate atoms; $\Omega_c=(\mathbf{e}_c\cdot \mathbf{p}_{43})\mathcal{E}_c/\hbar$ is the half Rabi frequency of the control field; $g_{p+}\equiv(\mathbf{e}_{p+}\cdot \mathbf{p}_{31})\mathcal{E}_p/\hbar$ [$g_{p-}\equiv(\mathbf{e}_{p-}\cdot \mathbf{p}_{32})\mathcal{E}_p/\hbar$] is the single-photon half Rabi frequency denoting the dipole coupling between the $\sigma^+$ ($\sigma^-$) component of the probe field and the atomic transition $|1\rangle\leftrightarrow|3\rangle$ ($|2\rangle\leftrightarrow|3\rangle$). Here,
$\mathbf{p}_{\alpha\beta}$ is the electric dipole matrix element associated with the atomic transition from $|\beta\rangle$ to $|\alpha\rangle$,  $g_{p+}\approx g_{p-}=g_{p}$ due to symmetry of the level configuration of the double Rydberg-EIT. In addition, we have defined
$\hat{S}_{\alpha\beta}\equiv |\beta\rangle\langle\alpha|\exp[i(k_{\beta}-k_{\alpha})z-i(\omega_\beta
-\omega_\alpha+\Delta_\beta-\Delta_\alpha)t]$
as atomic transition operators related to the states $|\alpha\rangle$ and $|\beta\rangle$ $(\alpha,\beta = 1$-4),
with $k_1=0$, $k_2=k_{p+}-k_{p-}=0$, $k_3=k_{p+}$, $k_4=k_{p+}+k_c$, $\omega_{\alpha}=E_{\alpha}/\hbar$ ($E_{\alpha}$ being the eigenenergy of the atomic state $|\alpha \rangle$)~\cite{Mu2021}.
$\hat{S}_{\alpha\beta}$ obey the commutation relation
\begin{eqnarray}\label{S commutation relation}
&& [\hat{S}_{\alpha\beta}(z,t),\hat{S}_{\mu\nu}(z^{\prime},t)]
\nonumber\\
&& =\frac{L}{N} \delta(z-z^{\prime})[\delta_{\alpha\nu}\hat{S}_{\mu \beta}(z,t)-\delta_{\mu \beta}\hat{S}_{\alpha\nu}(z,t)],\nonumber
\end{eqnarray}
with $N$ the total atomic number of the system. Note that, as in Refs.~\cite{Baur2014PRL,Gorniaczyk2014,Tiarks2014PRL,Gorniaczyk2016,Li2015,
Murray2016,Tiarks2016,Tiarks2019}, when writing (\ref{Hami2}) and (\ref{Hami3}) we have assumed $\rho_a$ is small and hence the Rydberg-Rydberg interaction between the atoms excited in the Rydberg state $|4\rangle$ is negligible.

The Zeeman effect induced by the magnetic field $B$ makes the levels $|1\rangle$ and $|2\rangle$ (which are degenerate when $B =0$) produce splitting $\Delta E=\mu_{\mathrm{B}} g_F^{\alpha} m_F^{\alpha} B$. Here $\mu_{\mathrm{B}}$, $g_F^{\alpha}$, and $m_F^{\alpha}$ are Bohr magneton, gyromagnetic factor, and magnetic quantum number of the atomic state $|\alpha\rangle$, respectively.
Therefore, we have $\Delta_2= -\Delta_1=(E_2-E_1)/ 2\hbar=\mu_{21} B/2\hbar$, $\Delta_3=\omega_p-(E_3-E_1)/\hbar$, and $\Delta_{4}=\omega_{p}+\omega_{c}-(E_4-E_1)/\hbar-\mu_{41} B/\hbar$, with $\mu_{\alpha \beta}= \mu_{\mathrm{B}} (m_F^{\alpha} g_F^{\alpha}- m_F^{\beta} g_F^{\beta})$.
The derivation of the effective Hamiltonian (\ref{Hami}) is similar to that given in the Appendix A of Ref.~\cite{Mu2021}.

As indicated above, we are interested in the case of a single gate photon stored in the gate atom located at the position $z_g$. Thus the gate-atom density is given by $\rho_g(z'_g)=\delta(z^{\prime}_g-z_g)$, and  $\hat{S}_{33}(z_g,t)\approx \hat{I}$ ($\hat{I}$ is unit matrix). For simplicity, we assume  $\rho_{a}$ is a constant, given by $\rho_a= N/L$.
Then the Hamiltonian $\hat{H}_{\rm AG}$ is reduced to the form
$\hat{H}_{\rm AG}=\rho_{a}\int_{-\infty}^{+\infty} dz\,\hbar \Delta_{d}(z){\hat S}_{44}(z,t)$, with
\begin{eqnarray}\label{Delta0}
\Delta_{d}(z)= \int_{-\infty}^{+\infty}dz_{g}' \rho_g(z'_g) \frac{-C_6}{\left|z_{g}'-z\right|^{6}}
=-\frac{C_6}{\left|z_g-z\right|^{6}},
\end{eqnarray}
As a result, $\Delta_{d}(z)$ behaves as a position-dependent detuning, which will
contribute an external optical potential (i.e. Rydberg defect potential) for the scattering of the incident probe photon polarization qubit (see Sec.~\ref{Sec3} below).
The position of the gate atom is assumed to locate at the middle of the atomic gas (i.e. $z_g=L/2$).

The time evolution of the atoms in the system is governed by the Heisenberg equation of motion
\begin{equation}\label{HL}
i\frac{\partial}{\partial t}{\hat S}_{\alpha\beta}=\left[{\hat S}_{\alpha\beta},\frac{\hat H}{\hbar}\right]+i{\hat{\cal L}}({\hat S}_{\alpha\beta})+i{\hat F}_{\alpha\beta}.
\end{equation}
Here the term ${\hat{\cal L}}({\hat S}_{\alpha\beta})$ describes the dissipation of ${\hat S}_{\alpha\beta}$ due to spontaneous emission and dephasing, ${\hat F}_{\alpha\beta}$ are $\delta$-correlated Langevin noise operators describing the fluctuations associated with the dissipations ${\hat{\cal L}}({\hat S}_{\alpha\beta})$. Explicit expression of Eq.~(\ref{HL}) are presented in Appendix~\ref{app1}. For simplicity, in the present work the dynamics of the gate atom is not considered, which is approximately valid because the lifetime the Rydberg state is quite
long~\cite{Baur2014PRL,Gorniaczyk2014,Tiarks2014PRL,Gorniaczyk2016,Tiarks2016,Tiarks2019}.

The evolution of the probe field is controlled by the Maxwell equation
$\partial^2\hat{\bf E}_{p}/\partial z^2-(1/c^2)\partial^2{\hat{\bf E}}_{p}/\partial t^2=(1/\varepsilon_{0}c^2)\partial^2{\hat{\bf P}}_{p}/\partial t^2$,
with ${\hat{\bf P}}_{p}\equiv
{\cal N }_{a}({\bf p}_{31}{\hat S}_{31}+{\bf p}_{32}{\hat S}_{32})e^{i(k_{p}z-\omega_{p}t)}+{\rm h.c.}$ the polarization intensity, ${\bf p}_{31}$  (${\bf p}_{32}$)  the electric dipole matrix element related to the transition from $|3\rangle$ to $|1\rangle$  ($|3\rangle$ to $|2\rangle$), and ${\cal N}_{a}\equiv N/V=\rho_a/A_0$ the volume atomic density.
Under slowly-varying approximation, the Maxwell equation is reduced to
\begin{equation}\label{Maxwell2}
\begin{aligned}
& i\left(\frac{\partial}{\partial z}+\frac{1}{c}\frac{\partial}{\partial t}\right){\hat E}_{p+}
+\frac{g_{p+}^{\ast}N}{c}{\hat S}_{31}=0,\\
& i\left(\frac{\partial}{\partial z}+\frac{1}{c}\frac{\partial}{\partial t}\right){\hat E}_{p-}
+\frac{g_{p-}^{\ast}N}{c}{\hat S}_{32}=0.
\end{aligned}
\end{equation}

The physical model described above is valid for many alkali-metal atomic gases, such as $^{85}$Rb, $^{87}$Rb, and $^{88}$Sr. In numerical calculations given in the following, we shall take cold $^{85}$Rb gas as an example. The atomic levels for realizing the double Rydberg-EIT are selected to be $|1\rangle=\left|5^2 S_{1 / 2}, F=3, m_F=-1\right\rangle$, $|2\rangle=\left|5^2 S_{1 / 2}, F=3, m_F=1\right\rangle$, $|3\rangle=\left|5^2 P_{3 / 2}, F=4, m_F=0\right\rangle$, and $|4\rangle=\left|68 S_{1 / 2}\right\rangle$. For $n=n'=68$ ($n$ and $n'$ are principal quantum numbers of the Rydberg states $|4\rangle$ and $|3\rangle_g$, respectively), the van der Waals dispersion parameter reads $C_{6}=-2\pi\times625.6\,{\rm GHz}\cdot\mu {\rm m}^6$
(i.e., the Rydberg-Rydberg interaction is repulsive). Other system parameters are given by $\Gamma_{12}=\Gamma_{21}=2\pi\times0.0016\,{\rm MHz}$, $\Gamma_{3}=2\pi\times6.06\,{\rm MHz}$, $\Gamma_{4}=2\pi\times0.02\,{\rm MHz}$, and  $\Gamma_{13}=\Gamma_{23}=\Gamma_{3}/2$.

For the $D2$ line of $^{85}$Rb atoms, the gyromagnetic factor of the two lower levels is $g_F = 1/3$.  Due to the symmetry of the lower energy level shifts induced by the magnetic field $B$, we have
\begin{equation}\label{Delta2}
 \Delta_1=-\Delta_2=-\frac{\mu_B B}{3 \hbar}.
\end{equation}
We stress that, due to the choice of magnetic quantum numbers and the linear polarization of the control field, the levels $|3\rangle$ and $|4\rangle$ are not sensitive to the applied magnetic field. Therefore, the dependence on $B$ for $\Delta_{3}$ and $\Delta_{4}$ is negligible.

\subsection{Envelope equations of the two-component probe field}\label{Sec23}
To study the propagation of the probe field under the action of the gate photon, we must solve the HM equations (\ref{HL}) and (\ref{Maxwell2}).
Because the probe field under consideration is at a single-photon level,  nonlinear terms in the HM equations are negligible. By employing Fourier transformation and eliminating atomic variables, we obtain the following linear envelope equations describing the dynamics of the two polarization components of the probe field in frequency space:
\begin{align}\label{RLSE}
\left[i\frac{\partial}{\partial z}+K_{j}(z,\omega)\right]{\tilde{\hat E}}_{pj}(z,\omega)=i{\tilde{\hat{{\cal F}}}}_{pj}(z,\omega),
\end{align}
where $j=+,-$, and
\begin{subequations}
\begin{eqnarray}\label{FT}
&& \tilde{\hat{E}}_{pj}(z, \omega)=\frac{1}{\sqrt{2 \pi}} \int_{-\infty}^{\infty} d t \hat{E}_{pj}(z, t) e^{-i \omega t}, \label{FT1}\\
&& K_{+}(z,\omega)=\frac{\omega}{c}+\frac{|g_{p}|^{2}N}{2c}\frac{[\omega+d_{41}
-\Delta_{d}(z)]}{D_{1}(\omega)},\label{FT2}\\
&& K_{-}(z,\omega)=\frac{\omega}{c}+\frac{|g_{p}|^{2}N}{2c}
\frac{[\omega+d_{42}-\Delta_{d}(z)]}{D_{2}(\omega)}. \label{FT3}
\end{eqnarray}
\end{subequations}
with $D_{\alpha}(\omega)=|\Omega_{c}|^2-(\omega+d_{3\alpha})
[\omega+d_{4\alpha}-\Delta_{d}(z)]$ ($\alpha = 1, 2$).
Here
$d_{\alpha\beta}=\Delta_{\alpha}-\Delta_{\beta}+i\gamma_{\alpha\beta}$
($\alpha\neq \beta)$, $\gamma_{\alpha\beta}\equiv(\Gamma_\alpha+\Gamma_\beta)/2+\gamma_{\alpha\beta}^{\rm dep}$, and $\Gamma_\beta\equiv\sum_{\alpha<\beta}\Gamma_{\alpha\beta}$. $\Gamma_{\alpha\beta}$ is the decay rate of the spontaneous emission from the state $|\beta\rangle$ to the state $|\alpha\rangle$, $\gamma_{\alpha\beta}^{\rm dep}$ is the dephasing rate between $|\alpha\rangle$ and $|\beta\rangle$.
Quantities $K_{+}(z,\omega)$ and $K_{-}(z,\omega)$ are linear
dispersion relations for the $\sigma^+$ and $\sigma^-$ polarization components, respectively. A detailed derivation on Eq.~(\ref{RLSE}) is presented in Appendix~\ref{app2}, with explicit expressions of Langevin noise terms ${\tilde{\hat{{\cal F}}}}_{pj}(z,\omega)$ given by Eqs.~(\ref{LBF1}) and (\ref{LBF2}).

Notice that when deriving Eq.~(\ref{RLSE}), we have, for simplicity, assumed that $\Delta_d (z)$ is a slowly-varying function of $z$. This allows to take $\Delta_d (z)$ be approximated as a constant during the Fourier transformation~\cite{noteapp2}.
Under the double EIT condition (i.e. $|\Omega_c|^2\gg \gamma_{3\alpha}\gamma_{4\alpha}$; $\alpha=1,2$), the Langevin noise terms ${\tilde{\hat{{\cal F}}}}_{pj}(z,\omega)$ in the envelope equations (\ref{RLSE}) are very small and hence can be neglected safely~\cite{Gorshkov2011PRL,Zhu2021,Zhu2022,Zhu2023}.

In the absence of the  control field (i.e. $\Omega_{c}=0$) and the gate atom [i.e. $\Delta_d(z)=0$], amplitudes of the two polarization components of the probe field (i.e. $\hat{E}_{p+}$ and $\hat{E}_{p-}$) behave in the way of exponential decay with the form  $\exp(-OD)$ when passing through the atomic medium, with  $OD=|g_{p}|^{2}N L/(2 c \gamma_{31})$ the optical depth of the atomic gas
(which describes the effective coupling strength between the probe field and the atoms).
The application of the control field (i.e. $\Omega_{c}\neq 0$) induces de-construction quantum interference effects for atomic transition paths, so that transparent windows on the absorption spectra of the two polarization components will open, resulting in the occurrence of the double Rydberg-EIT phenomenon in the system. For completeness, a detailed discussion on the double Rydberg-EIT in the absence of the gate atom is given in
Appendix~\ref{app3}.

\section{Switch and phase shifts of photon polarization qubit}\label{Sec3}

\subsection{Rydberg-defect potential for manipulating photon polarization qubit}\label{Sec31}
As illustrated above, the existence of the gate atom contributes the position-dependent detuning $\Delta_d(z)$. In fact, this position-dependent detuning can induce a Rydberg-defect potential for the propagation of the probe pulse. To see this clearly, we write Eq.~(\ref{RLSE}) into the following form
\begin{eqnarray}\label{RLSEA}
&& i\hbar\frac{\partial}{\partial \tau} \tilde{\hat E}_{pj}(z,\omega)=V_{j}(z,\omega)\tilde{\hat E}_{pj}(z,\omega),
\end{eqnarray}
after neglecting the small Langevin noise terms. Here $\tau\equiv ct$ and $V_{j}(z,\omega)\equiv -\hbar cK_{j}(z,\omega)$ ($j=+,-$). One sees that $V_{+}(z,\omega)$ and $V_{-}(z,\omega)$ play roles of external potentials for the $\sigma^+$ and $\sigma^-$ polarization components, respectively.
It is the $z$-dependence of $V_{\pm}$ that contributes the Rydberg defect potential to the propagation of the probe pulse, and hence induces switch behavior and phase shifts for the photon polarization qubit.

For simplicity, here we give only a detailed discussion on $V_{\pm}(z,\omega)$ near the center point of the EIT transparency windows (i.e. $\omega=0$). According to (\ref{FT2}) and (\ref{FT3}), we have $V_{\pm}(z,0)\equiv V_{\pm}(z) = {\rm Re}[V_{\pm}(z)]+i\,{\rm Im}[V_{\pm}(z)]$, which means that the Rydberg defect potential has real and imaginary parts.
Detailed expressions of ${\rm Re}[V_{\pm}(z)]$ and ${\rm Im}[V_{\pm}(z)]$ are presented in Appendix~\ref{app4}.

To simplify the expressions of ${\rm Re}[V_{\pm}(z)]$ and ${\rm Im}[V_{\pm}(z)]$, we note that
the decay rates $\gamma_{41}$ and $\gamma_{42}$ can be approximated to be zero because the Rydberg state $|4\rangle$ has long lifetime; moreover,
for weak magnetic field $B$, detunings $\Delta_{2}=-\Delta_{1}\ll \Delta_{d}(z)$. Under such a consideration, ${\rm Re}[V_{\pm}(z)]$ and ${\rm Im}[V_{\pm}(z)]$ can be reduced to the simple forms
\begin{subequations}\label{VA1}
\begin{align}
& {\rm Re}[V_{+}(z)]\approx\frac{{\cal N}_{a}\omega_{p}|{\bf e}_{p+}\cdot{\bf p}_{31}|^2}{4\varepsilon_{0}}\nonumber\\
&\hspace{1.7cm}\times \frac{\Delta_{d}(z)[|\Omega_{c}|^2
+(\Delta_{3}-\Delta_{1})\Delta_{d}(z)]}{\left||\Omega_{c}|^2+d_{31}\Delta_{d}(z)\right|^2},
\label{VA11}\\
& {\rm Re}[V_{-}(z)]\approx\frac{{\cal N}_{a}\omega_{p}|{\bf e}_{p-}\cdot{\bf p}_{32}|^2}{4\varepsilon_{0}}\nonumber\\
&\hspace{1.7cm}\times
\frac{\Delta_{d}(z)[|\Omega_{c}|^2
+(\Delta_{3}-\Delta_{2})\Delta_{d}(z)]}{\left||\Omega_{c}|^2+d_{32}\Delta_{d}(z)\right|^2},
\label{VA12}\\
& {\rm Im}[V_{+}(z)]\approx-\frac{{\cal N}_{a}\omega_{p}|{\bf e}_{p+}\cdot{\bf p}_{31}|^2}{4\varepsilon_{0}}\frac{\gamma_{31}|\Delta_{d}(z)|^2}{\left||\Omega_{c}|^2
+d_{31}\Delta_{d}(z)\right|^2},\label{VA13}\\
& {\rm Im}[V_{-}(z)]\approx-\frac{{\cal N}_{a}\omega_{p}|{\bf e}_{p-}\cdot{\bf p}_{32}|^2}{4\varepsilon_{0}}\frac{\gamma_{32}|\Delta_{d}(z)|^2}{\left||\Omega_{c}|^2
+d_{32}\Delta_{d}(z)\right|^2}.\label{VA14}
\end{align}
\end{subequations}
From these expressions, we see that the Rydberg-defect potential $V_{\pm}(z)={\rm Re}[V_{\pm}(z)]+i{\rm Im}[V_{\pm}(z)]$ are proportional to the position-dependent detuning $\Delta_{d}(z)$.

\begin{figure*}
\centering
\includegraphics[width=1.9\columnwidth]{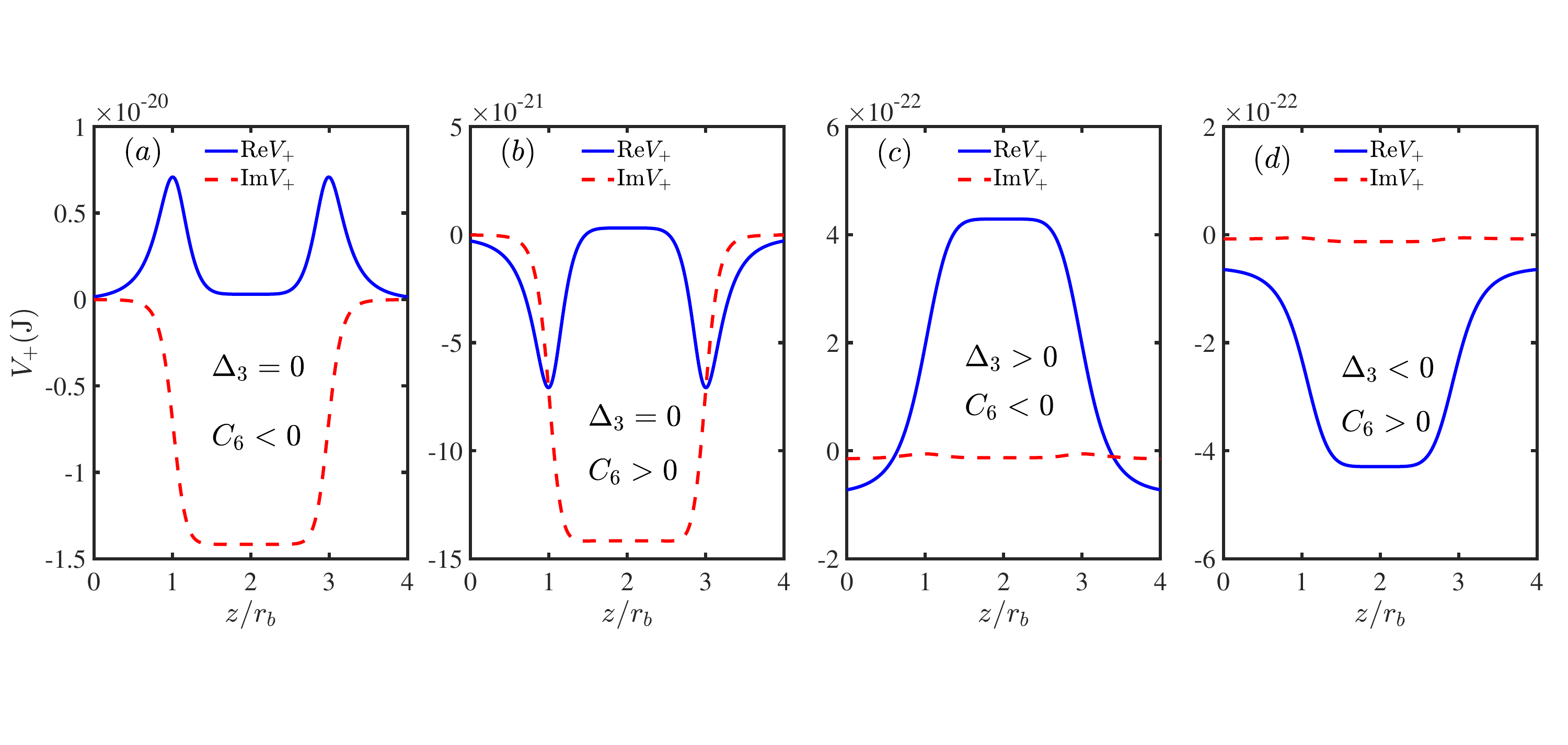}
\caption{{\footnotesize Rydberg-defect potential $V_{+}(z)={\rm Re}[V_{+}(z)]+i{\rm Im}[V_{+}(z)]$.
(a)~Solid  blue line and dashed red line are respectively the real part ${\rm Re}[V_{+}(z)]$  and the imaginary part ${\rm Im}[V_{+}(z)]$ as a function of $z/r_{b}$, by taking $\Delta_{3}=\Delta_{4}=0$, $\Omega_c=2\pi\times6.37$\,MHz, ${\cal N}_{a}=3\times10^{12}\,{\rm cm}^{-3}$,  $C_{6}=-2\pi\times 625.6\,{\rm GHz}\cdot\mu {\rm m}^6$.
The system works in a dissipation regime for the propagation of the probe pulse, useful for designing qubit switches.
(b)~The same as (a), but for $C_{6}=2\pi\times 625.6\,{\rm GHz}\cdot\mu {\rm m}^6$.
(c)~The same as (a), but with $\Delta_{3}=2\pi\times100\,{\rm MHz}$.
The system works in a dispersion regime for the propagation of the probe pulse, useful for realizing large phase shifts for photon qubits.
(d)~The same as (a), but with $\Delta_{3}=-2\pi\times100\,{\rm MHz}$ and $C_{6}=2\pi\times 625.6\,{\rm GHz}\cdot\mu {\rm m}^6$.
Due to the symmetry of the excitation configuration of the double Rydberg-EIT, $V_{-}(z) \approx V_{+}(z)$, thus not shown.
}}
\label{Fig2}
\end{figure*}
Various profiles of the Rydberg defect potential as functions of $z/r_{b}$ for different system parameters are given in Fig.~\ref{Fig2}. Solid  blue line and dashed red line in Fig.~\ref{Fig2}(a) are respectively for the real part ${\rm Re}[V_{+}(z)]$ and the imaginary part ${\rm Im}[V_{+}(z)]$, by taking $\Delta_{3}=\Delta_{4}=0$, $\Omega_c=2\pi\times6.37$\,MHz, ${\cal N}_{a}=3\times10^{12}\,{\rm cm}^{-3}$, $C_{6}=-2\pi\times 625.6\,{\rm GHz}\cdot\mu {\rm m}^6$. From the figure, we see that $|{\rm Im}[V_{+}(z)]|$ is much larger than $|{\rm Re}[V_{+}(z)]|$. This is due to the selection of vanishing single-photon detuning (i.e. $\Delta_{3}=0$), which makes the system work in a dissipation regime for the propagation of the probe field,
useful for designing qubit switches (see Sec.~\ref{Sec32} below).
The result shown by Fig.~\ref{Fig2}(b) is obtained by using $C_{6}=2\pi\times 625.6\,{\rm GHz}\cdot\mu {\rm m}^6$ [other parameters are the same as those  in Fig.~\ref{Fig2}(a)]. In this case the system works still in dissipation propagation regime (i.e. $|{\rm Im}[V_{+}(z)]|\gg |{\rm Re}[V_{+}(z)]|$).

Plotted in Fig.~\ref{Fig2}(c) are profiles of ${\rm Re}[V_{+}(z)]$ and ${\rm Im}[V_{+}(z)]$ by selecting a large and positive single-photon detuning ($\Delta_{3}=2\pi\times100\,{\rm MHz}$), with other system parameters the same as those  in Fig.~\ref{Fig2}(a). One sees that in this situation the real part of the potential is much larger than its imaginary part (i.e. $|{\rm Re}[V_{+}(z)]|\gg|{\rm Im}[V_{+}(z)]|$). This fact tells us that the selection of positive and large single-photon detuning $\Delta_{3}$ can make the system work in a dispersive regime and ${\rm Re}[V_{+}(z)]$ has a shape of single barrier (repulsive), useful for realizing large phase shifts for photon qubits (see Sec.~\ref{Sec33} below).
The result given by Fig.~\ref{Fig2}(d) is obtained by using $\Delta_{3}=-2\pi\times100\,{\rm MHz}$ and $C_{6}=2\pi\times 625.6\,{\rm GHz}\cdot\mu {\rm m}^6$, with other parameters are the same as those in Fig.~\ref{Fig2}(a). In this case, the system works still in dispersion regime, but ${\rm Re}[V_{+}(z)]$ displays a shape of single well. This means that the Rydberg defect potential of this case can be used to trap the photon polarization qubit, which is interesting, but will not be discussed in the present work.

Note that, by inspecting the symmetry of the excitation configuration of the double Rydberg-EIT [see the left part of Fig.~\ref{Fig1}(a)], we have $V_{-}(z) \approx V_{+}(z)$. Thus the profile of the Rydberg-defect potential $V_{-}(z)$ is basically the same as that of $V_{+}(z)$. This point can be seen clearly from the expressions given by (\ref{VA11})-(\ref{VA14}).

Based on the above analysis, we see that the dispersion coefficient of  Rydberg-Rydberg interaction $C_6$ and the single-photon detuning $\Delta_{3}$ are two important parameters for controlling the property of the Rydberg defect potential. Based on such results, we can realize various Rydberg defect potentials and hence can actively manipulate the behavior of the incident photon polarization qubits. In the following discussions, we consider only two cases for $C_6<0$ [i.e. the Rydberg defect potentials of the forms shown in Fig.~\ref{Fig2}(a) and Fig.~\ref{Fig2}(c)].

\subsection{Switch of the photon polarization qubit in dissipation regime}\label{Sec32}

We now explore the possibility of new type of photon switch in the system. Single photon switches are optical devices for controlling the transmission of target photons through the application only a single gate photon. They are key devices for all-optical quantum information processing~\cite{Miller2010}. One of techniques for building single photon switches is the use of the dissipative optical nonlinearity via Rydberg-EIT. In the past few years, the possibility for realizing such switches for target photons with one polarization component have been demonstrated
experimentally~\cite{Baur2014PRL,Gorniaczyk2014,Tiarks2014PRL,Gorniaczyk2016}.
Here, we show that the model proposed above can be used to realize another type of single photon switch, which is for the single photon with two polarization components (i.e. photon polarization qubit switch). The basic idea of the scheme is the following. First, a single gate photon is stored in the Rydberg state $|3\rangle_g$ [as shown in the right part of Fig~\ref{Fig1}(a)], which provides the Rydberg defect potential discussed in the last subsection. Second, a probe photon qubit (as a target photon) with $\sigma^+$ and $\sigma^-$ polarization components is incident into the atomic gas working in the dissipation regime of the double Rydberg-EIT (realized by taking zero single-photon detuning, i.e. $\Delta_3=0$), for which the imaginary part of the Rydberg defect potential is much bigger than its real part [see Fig.~\ref{Fig2}(a)].  When the gate photon is absent, the photon polarization qubit would propagate in the atomic gas nearly without absorption [as schematically shown in the top of Fig~\ref{Fig1}(c)]; however, when the gate photon is present, the strong Rydberg-Rydberg interaction between the states $|4\rangle$ and $|3\rangle_g$ results in a Rydberg blockade effect (the breaking of the double Rydberg-EIT), and hence switches the atomic gas from highly transparent to strongly absorptive [as shown in the middle of Fig~\ref{Fig1}(c)].

To this end, we consider the dynamics of the two polarization components of the probe pulse in the presence of the Rydberg defect potential, which is controlled by the envelope equation (\ref{RLSE}).
By directly integrating Eq.~(\ref{RLSE}) from $0$ to $L$, we get the solution (in frequency domain):
\begin{equation}\label{PPRO2}
\tilde{{\hat E}}_{pj}(L,\omega)=\tilde{{\hat E}}_{pj}(0,\omega) \, \exp \left[ i\int_{0}^{L}  dz K_{j}(z,\omega) \right],
\end{equation}
with $j=+,-$. The solution in time domain can be obtained by using inverse Fourier transformation, which reads
\begin{eqnarray}\label{PPRO3}
&&{\hat E}_{pj}(L,t)=\int_{-\infty}^{+\infty} d\omega  \tilde{{\hat E}}_{pj}(0,\omega) \nonumber\\
&&\hspace{1.7cm}\times \exp \left[ i\int_{0}^{L}  dz K_{j}(z,\omega) -i \omega \left(t-\frac{L}{c}\right)\right].\nonumber\\
\end{eqnarray}
Here $\tilde{{\hat E}}_{pj}(0,\omega)$ is the Fourier transform of ${\hat E}_{pj}(z,t)$  at the input boundary $z=0$. Since the probe field is a pulse, $\tilde{{\hat E}}_{pj}(0,\omega)$ is narrow in $\omega$, and hence we can expand
$K_{j}(z,\omega)$ near $\omega=0$, i.e. $K_{j}(z,\omega) = K_{0j} + \omega K_{1j} +\cdots $, with $K_{0j}\equiv K_{j}(z,\omega)|_{\omega=0}$ and $K_{1j} \equiv (\partial K_{j}/\partial\omega)|_{\omega=0}$.
Then (\ref{PPRO3}) can be reduced to the following form
\begin{equation}\label{PPRO4}
{\hat E}_{pj}(L,t)\approx{\hat E}_{pj}(0,t-L^{\prime}/V_{gj}) e^{-\eta_{j}+i\phi_{j}},
\end{equation}
where
\begin{subequations}\label{PS0}
\begin{align}
\phi_{+}& =\frac{|g_{p}|^2 N }{c} {\rm Re} \left(\int_{0}^{L} dz\,  a_{31}(z)\right), \label{PS01}  \\
\phi_{-}& =\frac{|g_{p}|^2 N }{c} {\rm Re} \left(\int_{0}^{L} dz\,  a_{32}(z)\right),  \label{PS02}  \\
\eta_{+}& =\frac{|g_{p}|^2 N }{c} {\rm Im} \left(\int_{0}^{L} dz\,  a_{31}(z)\right),  \label{PS03}  \\
\eta_{-}& =\frac{|g_{p}|^2 N }{c} {\rm Im} \left(\int_{0}^{L} dz\,  a_{32}(z)\right),  \label{PS04}  \\
L^{\prime}& = L-2r_b\approx L, \label{PS05}
\end{align}
\end{subequations}
with
\begin{subequations}\label{it1}
\begin{align}
a_{31}(z)&=\frac{d_{41}-\Delta_{d}(z)}{2\{|\Omega_{c}|^2-d_{31}[d_{41}
-\Delta_{d}(z)]\}},\label{it11}\\
a_{32}(z)&=\frac{d_{42}-\Delta_{d}(z)}{2\{|\Omega_{c}|^2-d_{32}[d_{42}
-\Delta_{d}(z)]\}}.\label{it12}
\end{align}
\end{subequations}
In these expressions, $L^{\prime}$ is the reduced medium length due to
the existence of the Rydberg blockade;
$V_{gj}=[K_{1j}]^{-1} \equiv [(\partial K_{j}/\partial\omega)|_{\omega=0}]^{-1}$ is the group velocity of the $j$th polarization component, which is a constant after the pulse passes over the gate atom. Using the system parameters given in Sec.~\ref{Sec21} and taking
${\cal N}_{a}=3\times10^{12}\,{\rm cm}^{-3}$ and $\Omega_c=2\pi\times6.37$MHz,
we obtain $V_{g-}\approx V_{g+}=6.46\times 10^{-7} c$ (i.e. the probe pulse is a slow-light qubit).

The dynamics of the incident photon polarization qubit under the action of the Rydberg defect potential is characterized by key quantities $\eta_{j}$ and $\phi_{j}$ ($j=+, -$) in the solution (\ref{PPRO4}), which describe the amplitude attenuations and phase shifts of the two polarization components after traversing the gate atom, respectively. To demonstrate this, we first consider the switch behavior of the photon polarization qubit by assuming that the system works in the dissipative regime of the double Rydberg-EIT (i.e. $\Delta_{3}=0$).

\begin{figure}
\includegraphics[width=1.\columnwidth]{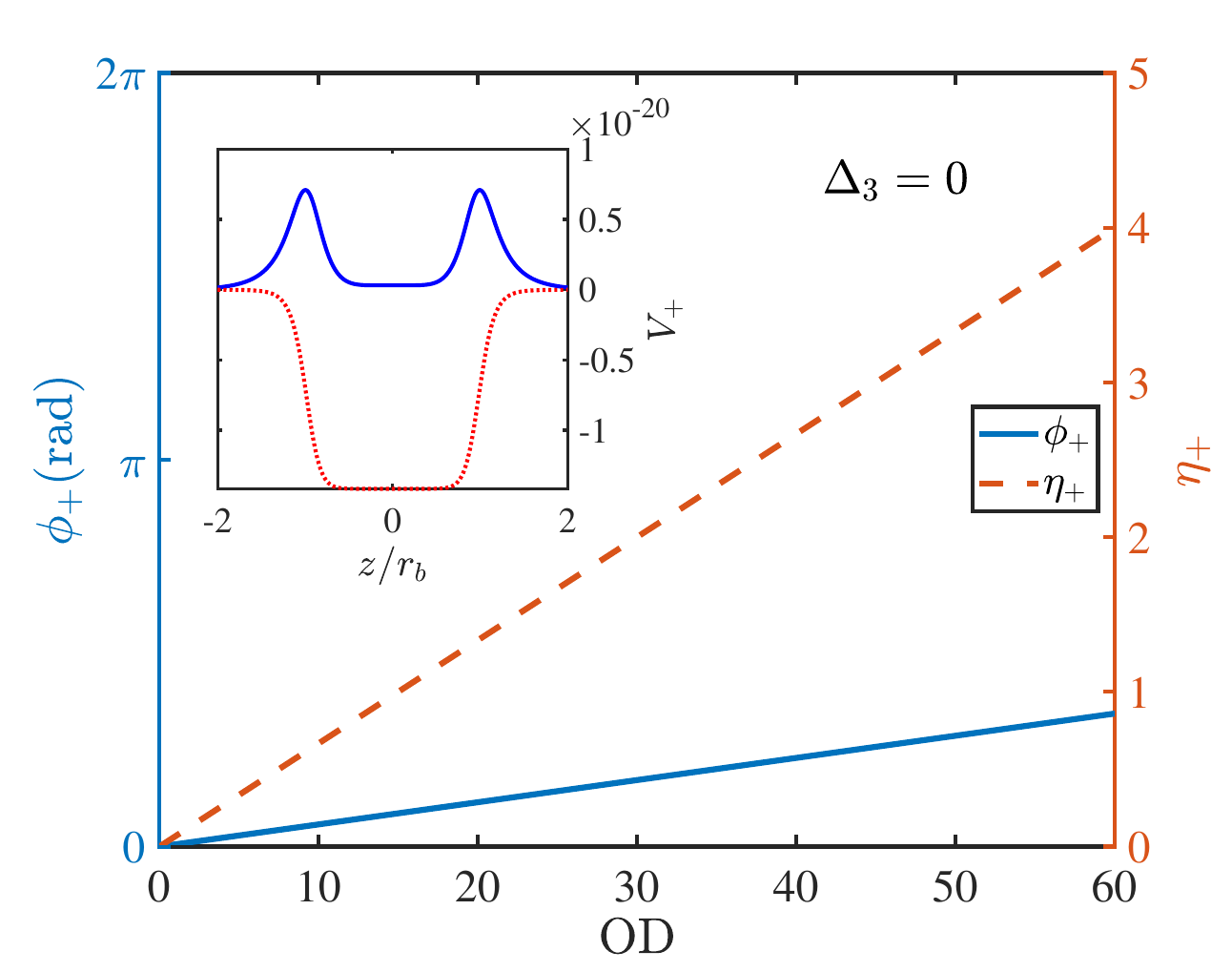}  
\caption{{\footnotesize
Switch of the photon polarization qubit in dissipation regime of the double Rydberg-EIT ($\Delta_{3}=0$). Dashed red line is the amplitude attenuation factor $\eta_{+}$  for the $\sigma^+$ polarization component, as a function of optical depth $OD\equiv|g_{p}|^{2}N L/(2c\gamma_{31})$.
$\eta_{+}$ reaches value 1 at ${\rm OD}\approx 15$; it grows rapidly as $OD$ is increased further. Phase shift $\phi_{+}$  in this regime is also shown by the solid blue line. Inset: the shape of the Rydberg defect potential $V_+$.
Because of the rapid exponential attenuation of the amplitude of the incident photon, the system acts as a well-behaved photon qubit switch.
Due to the excitation configuration symmetry of the double Rydberg-EIT, behaviors of the amplitude attenuation factor $\eta_{-}$  and the phase shift $\phi_{-}$  for the $\sigma^-$ component are similar to those of the $\sigma^+$ component, and hence not shown.
}}
\label{Fig3}
\end{figure}

Shown in Fig.~\ref{Fig3}(a)
is the numerical result for the qubit switch. The
dashed red line in the figure is the amplitude attenuation factor $\eta_{+}$ for the $\sigma^+$ polarization component [using the expression given by (\ref{PS03})], plotted as a function of optical depth $OD$. The phase shift $\phi_{+}$  in this regime is also displayed by the solid blue line.
The result is obtained for cold $^{85}$Rb atomic gas, by taking $B=1.5$\,{\rm G} (i.e., $\Delta_2=-\Delta_1=2\pi \times 0.7$\,MHz), $\Delta_{4}=0$, and other system parameters the same as in Sec.~\ref{Sec21}~\cite{note1}. The inset of the figure gives the shape of the Rydberg defect potential [the same as Fig.~\ref{Fig2}(a)]. Since in the double Rydberg-EIT there exists a configuration symmetry for excitation paths of the $\sigma^+$ and $\sigma^-$ polarization components, the amplitude attenuation factor $\eta_{-}$  and phase shift $\phi_{-}$
for the $\sigma^-$ component have respectively similar behavior as $\eta_{+}$ and $\phi_{+}$, and thus omitted here.

From the figure, we see that $\eta_{+}$ reaches value 1 when ${OD}\approx 15$, and it grows rapidly as $OD$ is increased further. This fact tells us that, because of the rapid exponential attenuation of the incident photon amplitude, the stored gate photon can act indeed as a well-behaved single-photon switch, which can significantly impede the transmission of the incident photon polarization qubit [as shown by the middle part of Fig.~\ref{Fig1}(c)].

\subsection{Phase shifts of the photon polarization qubit in dispersion regime}\label{Sec33}
We now turn to consider how to get large phase shifts for the photon polarization qubit. It is known that strong dispersive interaction between gate photon and target photon can be used to access significant phase shift for the target photon, which is also important for all-optical quantum information processing~\cite{Friedler2005,Chang2014,Gorshkov2011PRL,Tiarks2016,Tiarks2019,
Vaneecloo2022,Stolz2022,Murray2016adv,
Adams2020,Nielsen2000}. Here we show that large phase shifts for the two polarization components of the incident photon polarization qubit can be acquired
under the action of the gate photon if the system works in the dispersion regime of the double Rydberg-EIT.

The one-photon detuning $\Delta_3$ is a key parameter to control
the dissipation and dispersion behaviors of the system. When $|\Delta_3|\gg \gamma_{31}$, the system works in dispersion regime.
Notice that the general solution of the envelope equation (\ref{RLSE}), given by (\ref{PPRO4}), together with (\ref{PS0}) and (\ref{it1}), is valid for any value of the one-photon detuning $\Delta_3$. It thus can also be used to calculate the phase shifts $\phi_{j}$ and amplitude attenuation factors $\eta_{j}$ ($j=+,-$) of the photon polarization qubit for non-zero $\Delta_3$.

\begin{figure}
\includegraphics[width=1\columnwidth]{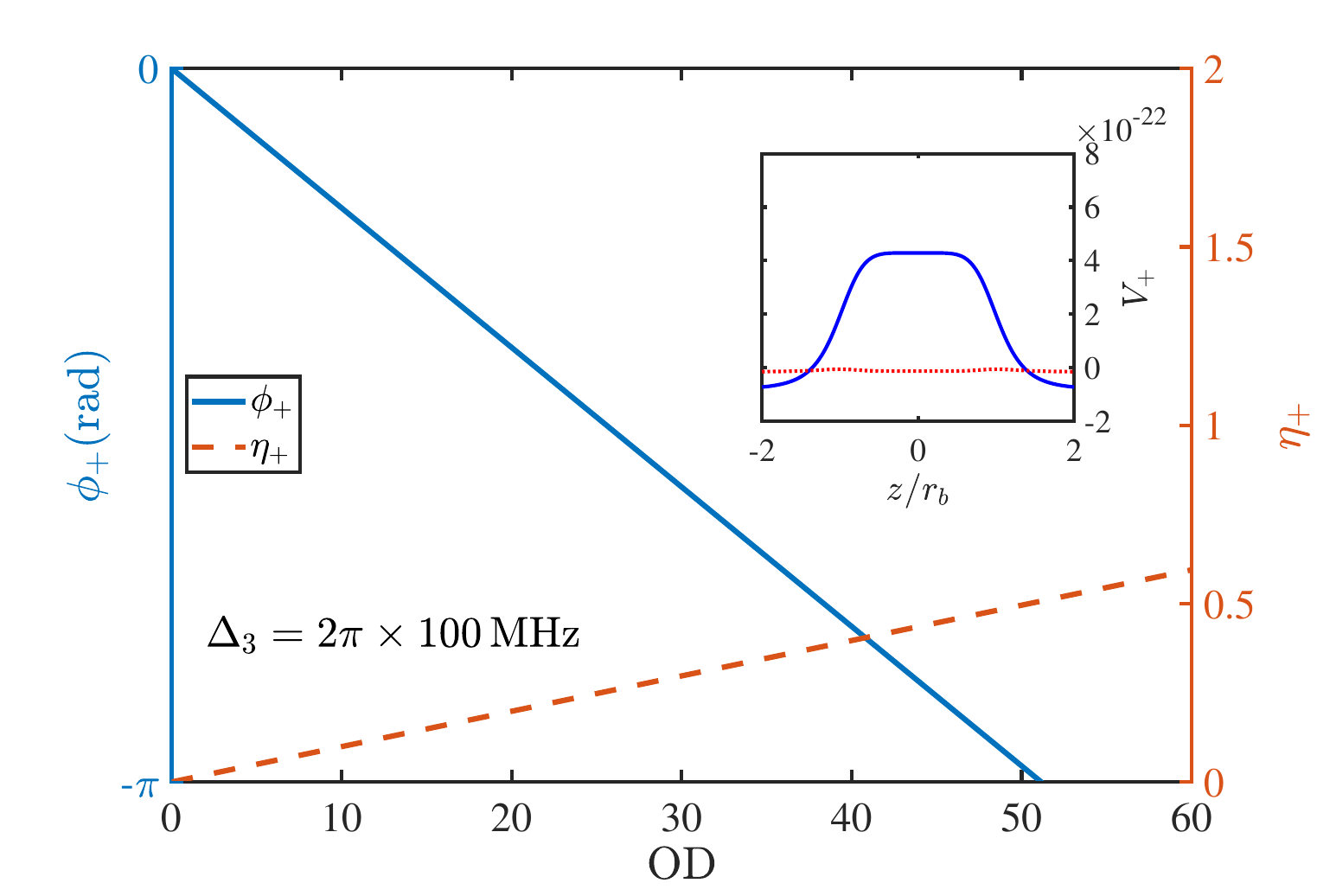}  
\caption{{\footnotesize
Phase shift of the photon polarization qubit in dispersion regime for the $\sigma^+$  component (with $\Delta_{3}=2\pi \times 100 ~ {\rm MHz}$, $B=1.5$\, {\rm G}).
Solid blue line is for $\phi_{+}$, as a function of optical depth $OD$.
$\phi_{+}$ reaches $-\pi$ radian for ${\rm OD}\approx 51$; it can be increased further as $OD$ increases. $\eta_{+}$  is also shown by the dashed red line, which is considered to be small in the range of ${\rm OD}\leq 51$ because the optical absorption is suppressed in this regime. Inset: the shape of the Rydberg defect potential $V_+$.
}}
\label{Fig4}
\end{figure}

\begin{figure}
\includegraphics[width=1\columnwidth]{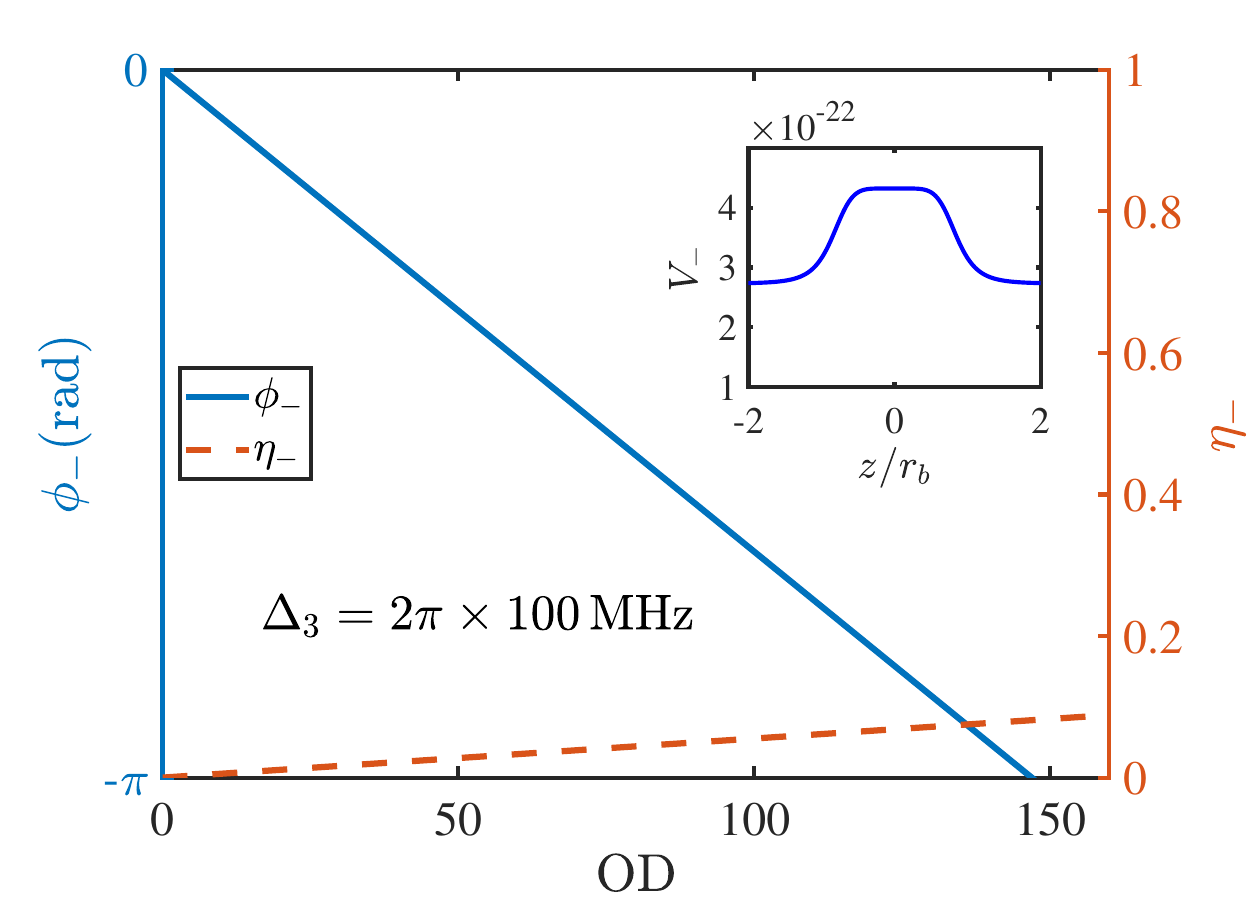}  
\caption{{\footnotesize
Phase shift of the photon polarization qubit in dispersion regime for the $\sigma^-$  component (with $\Delta_{3}=2\pi \times 100 ~ {\rm MHz}$, $B=1.5$\, {\rm G}).
Solid blue line is for $\phi_{-}$  as a function of optical depth $OD$.
$\phi_{-}$ reaches $-\pi$ radian for ${\rm OD}\approx 146$; it can be increased further as $OD$ increases. $\eta_{-}$  is also shown by the dashed red line, which is much smaller than $\eta_{+}$ because the optical absorption is greatly suppressed in this regime. Inset: the shape of the Rydberg defect potential $V_-$.
}}
\label{Fig5}
\end{figure}

Shown in Fig.~\ref{Fig4}
is the numerical result for the $\sigma^+$ polarization component in the dispersion regime ($\Delta_{3}=2\pi \times 100 ~ {\rm MHz}$).
The solid blue line in the figure is the phase shift $\phi_{+}$  as a function of optical depth $OD$. The result is obtained still for the cold $^{85}$Rb atomic gas, with $B=1.5$\, {\rm G} (i.e.,  $\Delta_2=-\Delta_1=2\pi \times 0.7$\,MHz), $\Delta_{4}=0$, and other parameters given in Sec.~\ref{Sec21}.
The inset of the figure gives the shape of the Rydberg defect potential [i.e. Fig.~\ref{Fig2}(c)].
Plotted in Fig.~\ref{Fig5}
is similar to Fig.~\ref{Fig4} but for $\phi_-$  and $\eta_{-}$ of the $\sigma^-$ component. For large $\Delta_3$ and non-zero $B$, the symmetry of the two Rydberg-EITs respectively for the $\sigma^+$ and $\sigma^-$ polarization components is broken, and hence the phase shift $\phi_{-}$ and the amplitude attenuation factor $\eta_{-}$  for the $\sigma^-$ polarization component have different behavior comparing with those of the $\sigma^+$ component.

From Fig.~\ref{Fig4} we see that $\phi_{+}$ reaches the value $-\pi$ radian for ${OD}\approx 51$, and it increases further as $OD$ is increased. The amplitude attenuation factors $\eta_{+}$  is also shown by the dashed red line; it is very small due the large $\Delta_3$, by which the photon absorption is greatly suppressed.
From Fig.~\ref{Fig5} one sees that although $\phi_{-}$ behaves similarly to $\phi_{+}$, a large optical depth (${OD}\approx 146$) is needed  to reach the value of $-\pi$. In addition, $\eta_{-}$ is much smaller than $\eta_{+}$, which can be seen by comparing  Fig.~\ref{Fig5} with Fig.~\ref{Fig4}.

Consequently, due to the existence of the Rydberg defect potential contributed by the stored gate photon, in the dispersion regime the two polarization components of the incident single-photon qubit can indeed acquire significant phase shifts with a very small attenuation of qubit amplitude.
However, these two polarization components display  different behaviors  due to large one-photon detuning $\Delta_3$ and the existence of non-zero magnetic field $B$.

\subsection{Propagation of qubit wavepacket}\label{Sec34}

To be more intuitive, we now present a study on the propagation of the photon polarization qubit when it passes through the Rydberg defect (gate atom) by using Schr\"odinger picture. In such an approach, the qubit can be described by a single-photon wavepacket with two polarization components.

Since the input probe pulse is in a single-photon qubit state, in the atomic medium  the photon state takes the form
\begin{widetext}
\begin{align}\label{SPWF}
&& |\Phi(t)\rangle=|\Phi_{+}(t)\rangle+|\Phi_{-}(t)\rangle=\int dz\left[\Phi_{+}(z,t){\hat E}_{p+}^{\dag}(z)+\Phi_{-}(z,t){\hat E}_{p-}^{\dag}(z)\right]|0\rangle.
\end{align}
\end{widetext}
Here $|0\rangle$ is electromagnetic vacuum, $\Phi_{j}(z,t)\equiv \langle0|{\hat E}_{pj}(z)|\Phi_{j}(t)\rangle$ is the effective wavefunction of the $j$th polarization component ($j=+,-$),
obeying the normalization condition $\int dz\left[|\Phi_{+}(z,t)|^2+|\Phi_{-}(z,t)|^2\right]=1$.

Based on Eq.~(\ref{RLSE}) and the above definition of one-photon state vector, it is easy to derive the equation
\begin{equation}\label{wavepac_trans}
  i\frac{\partial}{\partial z}\tilde{\Phi}_{j}(z,\omega)+K_{j}(z,\omega) \tilde{\Phi}_{j}(z,\omega)=0,
\end{equation}
where $\tilde{\Phi}_{j}(z,\omega)\equiv(1/\sqrt{2\pi}) \int_{-\infty}^{\infty} dt\, \Phi_j(z,t) e^{i \omega t}$
is the Fourier transform of ${\Phi}_{j}(z, t)$ ($j=+,-$).

We assume that the $j$-th component of the incident single-photon wavepacket has the Gaussian form
\begin{eqnarray}\label{Gaussian}
\Phi_{j}(0,t)=\sqrt{A_{j}}\sqrt{\frac{2 \sqrt{\ln (2)}}{t_0 \sqrt{\pi}}} \exp \left[-2 \ln (2) \frac{t^2}{t_0^2}\right],
\end{eqnarray}
where $A_{j}$ are amplitudes satisfying $A_{+}+A_{-}=1$, $t_0$ is the full width at half maximum (FWHM) of $|\Phi_{j}(0,t)|^2$. The Fourier transform of $\Phi_{j}(0,t)$
reads
\begin{eqnarray}\label{BC}
\tilde{\Phi}_{j}(0,\omega)=\sqrt{A_{j}}\sqrt{\frac{2\sqrt{\ln (2)}}{\omega_0 \sqrt{\pi}}} \exp \left[-2 \ln (2) \frac{\omega^2}{\omega_0^2}\right],
\end{eqnarray}
where $\omega_0=4 \ln (2)/ t_{0}$ is the FWHM of $|\tilde\Phi_{j}(0,\omega)|^2$.

\begin{figure}
\includegraphics[width=0.85\columnwidth]{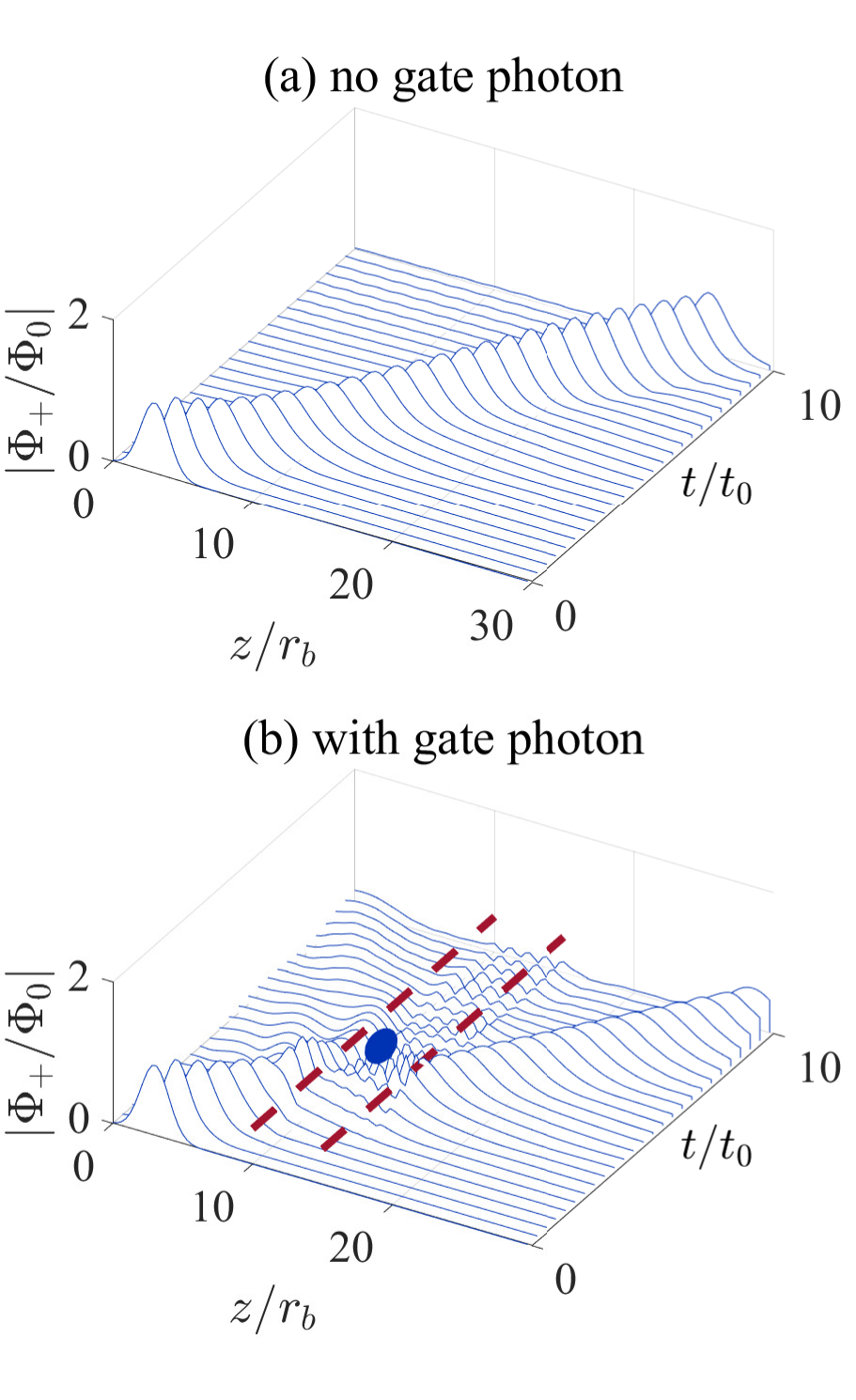}
\caption{{\footnotesize
Propagation of the wavepacket $\Phi_+$ of the $\sigma^+$ polarization component in the dispersion regime of the double Rydberg-EIT ($\Delta_3=2\pi\times100$\,MHz, $B=1.5$\, {\rm G}).
(a)~$\Phi_+$  as a function of time $t$ and position $z$ in the absence of the gate photon. $\Phi_{0}$ represents the initial amplitude of $\Phi_+$.
(b)~The same as (a) but for the presence of the gate photon. The width of the Rydberg defect potential is marked by two dashed red lines; the gate atom is denoted by the solid blue circle.
}}
\label{Fig6}
\end{figure}

By solving Eq.~(\ref{wavepac_trans}) under the boundary condition (\ref{BC}), we can obtain $\Phi_{j}(z,t)$ through the relation $\Phi_{j}(z,t)=(1/\sqrt{2\pi}) \int_{-\infty}^{\infty} d\omega\, \tilde{\Phi}_j(z,\omega) e^{-i \omega t}$.
Fig.~\ref{Fig6}(a)
shows $\Phi_+$ of the $\sigma^+$ polarization component as a function of time $t$ and spatial coordinate $z$ for the case of no gate photon, with $\Phi_{0}=\left[2 \sqrt{\ln (2)}A_{j}/(t_0 \sqrt{\pi})\right]^{1/2}$ representing the initial amplitude of $\Phi_+$. When plotting the figure, we have chosen $\Delta_3=2\pi\times100$\,MHz, $B=1.5$\, {\rm G}, $t_{0}=1\times10^{-7}$\,s, $A_+=1/2$, and ${\cal N}_{a}=3\times10^{12}\,{\rm cm}^{-3}$. We see that the wavepaeket propagates quite stably.
The reason is that, in the absence of the gate atom, the phase shift and attenuation of the wavepacket are nearly vanishing due to the EIT effect.

\begin{figure}
\includegraphics[width=0.85\columnwidth]{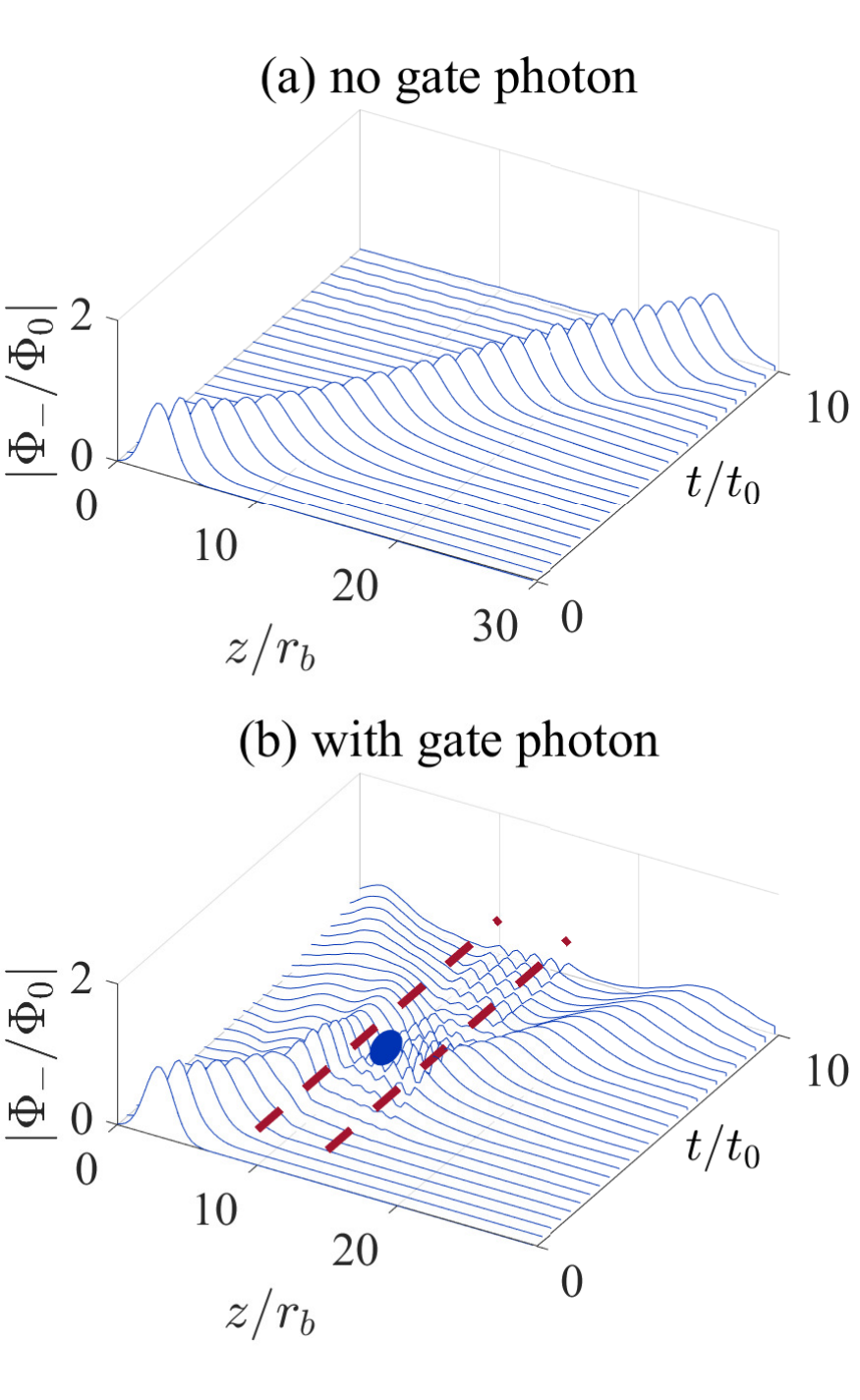}
\caption{{\footnotesize
Propagation of the wavepacket $\Phi_-$ of the $\sigma^-$ polarization component in the dispersion regime of the double Rydberg-EIT ($\Delta_3=2\pi\times100$\,MHz, $B=1.5$\, {\rm G}).
(a)~Wavefunction $\Phi_-$  as a function of time $t$ and position $z$ in the absence of the gate photon. $\Phi_{0}$ represents the initial amplitude of $\Phi_-$.
(b)~The same as (a) but for the presence of the gate photon. The width of the Rydberg defect potential is marked by two dashed red lines; the gate atom is denoted by the solid blue circle.
}}
\label{Fig7}
\end{figure}

Plotted in Fig.~\ref{Fig6}(b) is the wavefunction $\Phi_+$ as a function of $t$ and $z$ in the presence of the gate photon, with the system parameters the same as those used in Fig.~\ref{Fig6}(a). In the figure, the width of the Rydberg defect potential is indicated by the two dashed red lines, and the gate atom is denoted by the solid blue circle.
For comparision, Fig.~\ref{Fig7}
shows the propagation of the wavepacket $\Phi_-$ of the $\sigma^-$ polarization component. We see that in the presence of the gate atom $\Phi_-$ displays
a little different behavior comparing with that of $\Phi_+$.

If the photon polarization qubit is incident to the atomic gas at $(z,t)=(0,0)$,
the state vector of the probe field for this input state reads
$|\Phi_{+,{\rm in}}(0)\rangle =c_{+} |\sigma^{+} \rangle + c_{-} |\sigma^{-} \rangle$, with $c_{+}=\Phi_{+,{\rm in}}(0,0)$, $c_{-} =\Phi_{-,{\rm in}}(0,0)$,
$|\sigma^{+}\rangle={\hat E}^{\dag}_{p+}(0)|0\rangle$, and $|\sigma^{-}\rangle={\hat E}^{\dag}_{p-}(0)|0\rangle$.
Depending on whether 0 or 1 gate photon is stored, the
output qubit state (after passing through the atomic medium) is given by
\begin{subequations}\label{OS}
\begin{eqnarray}
&&\left|\Phi_{\text {out}, 0}\right\rangle \propto\left( c_+\left|\sigma^{+}\right\rangle+c_-\left|\sigma^{-}\right\rangle\right) \otimes|0\rangle_{g},\nonumber\\
\\
&&\left|\Phi_{\text {out}, 1}\right\rangle \propto\left(c_+ e^{-\eta_{+} } e^{i \phi_{+}}\left|\sigma^{+}\right\rangle+ c_- e^{-\eta_{-} } e^{i \phi_{-}}\left|\sigma^{-}\right\rangle\right) \nonumber\\
&& \hspace{1.7cm} \otimes|1\rangle_{g}.
\end{eqnarray}
\end{subequations}
Here, $|1\rangle_{g}$ ($|0\rangle_{g}$) is the Fock state with one gate photon (no gate photon) stored in the Rydberg state $|3\rangle_g$, $\eta_{j}$ and $\phi_{j}$ are respectively the amplitude attenuation factors and the phase shifts
for the $j$th polarization component ($j=+,-$), given by (\ref{PS01})-(\ref{PS04}).

\subsection{Magnetic-field-induced switching behavior of the photon polarization qubit}\label{Sec35}
How to detect weak magnetic fields is one of important topics in the study of precision measurements~\cite{Budker2013}. As final example, here we consider another possible applications of the strong interaction between the gate photon and the photon polarization qubit. We demonstrate that the present system can be used to design a new type of magnetometer that can be used to detect weak magnetic fields.

As indicated at the end of Sec.~\ref{Sec2}\,A, when the magnetic field ${\bf B}=(0,0,B)$ is applied to the system the Zeeman effect induced by the magnetic field makes the two degenerate levels $|1\rangle$ and $|2\rangle$  produce a level splitting proportional to $B$. Since $B$ is contained in the HM equations (\ref{HL}) and (\ref{Maxwell2}), solutions of
the amplitude attenuation factors $\eta_{j}$ and phase shifts $\phi_{j}$ ($j=+,-$), given by (\ref{PS01})-(\ref{PS04}), are also $B$-dependent. Hence, behaviors of the switch and phase shift of the photon polarization qubit can display a dependence on $B$.

\begin{figure}
\includegraphics[width=0.84\columnwidth]{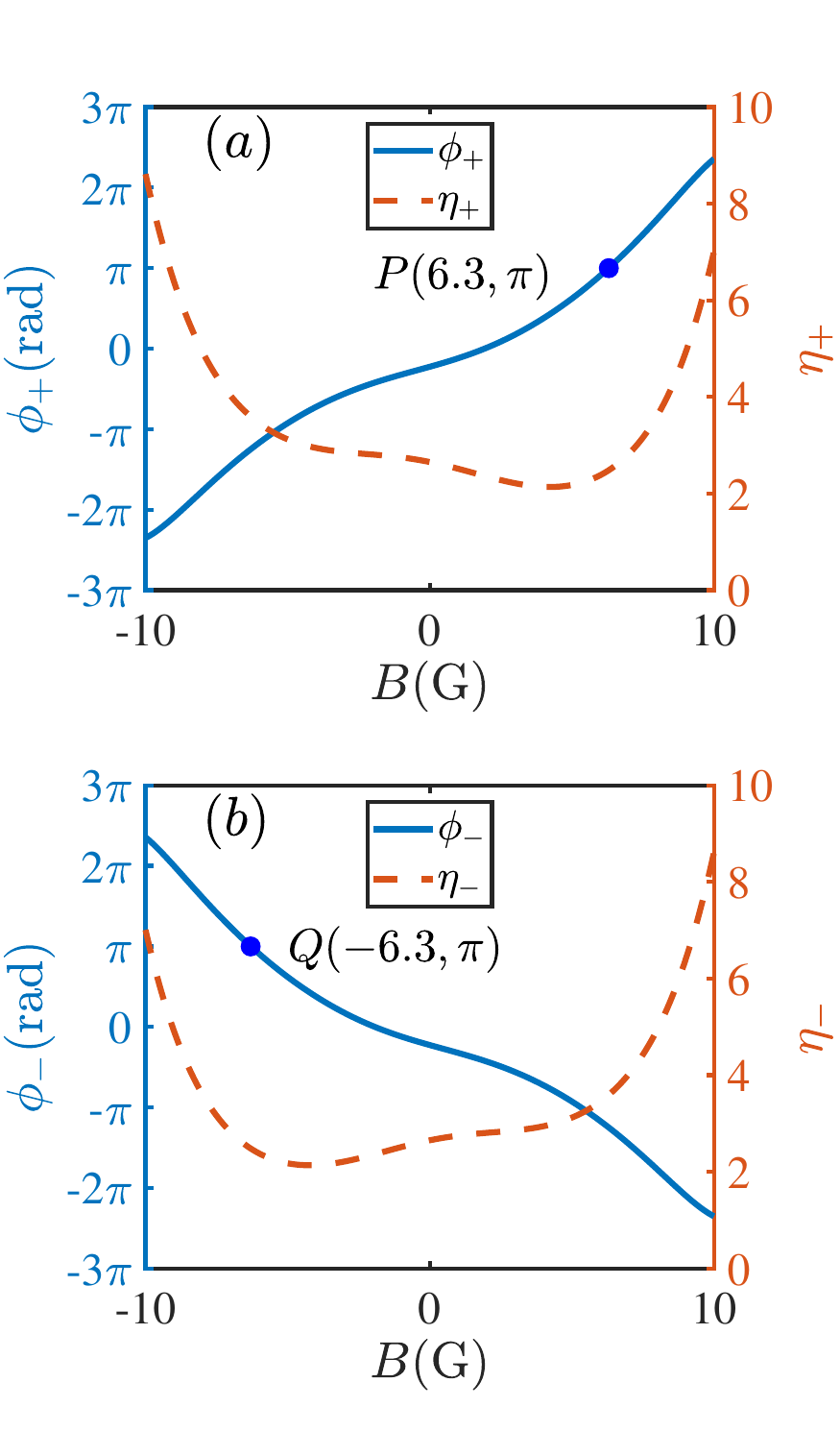}
\caption{{\footnotesize Magnetic-field-induced switching behavior of the photon polarization qubit.
(a)~ Amplitude attenuation factor $\eta_{+}$ (dashed red line) and phase shift $\phi_{+}$ (solid blue line) as functions of magnetic field $B$, for $\Delta_3=0$, ${\cal N}_{a}=3\times10^{12}\,{\rm cm}^{-3}$, and $L=50\,\mu{\rm m}$. $P\,(6.3,\pi)$ (solid circle) means the numerical point with $B=6.3$\,{\rm G} and $\phi_+=\pi$.
(b)~The same as (a) but for the amplitude attenuation factor $\eta_{-}$ (dashed red line) and phase shift $\phi_{-}$. $Q\,(-6.3,\pi)$ (solid circle) means the numerical point with $B=-6.3$\,{\rm G} and  $\phi_-=\pi$.
}}
\label{Fig8}
\end{figure}

Shown in Fig.~\ref{Fig8}(a) and Fig.~\ref{Fig8}(b)
are amplitude attenuation factors $\eta_{j}$ (dashed red lines) and phase shifts $\phi_{j}$ (solid blue lines) for the $j$-th polarization component ($j=+,-$), plotted as functions of the magnetic field $B$, by taking system parameters to be $\Delta_3=0$, ${\cal N}_{a}=3\times10^{12}\,{\rm cm}^{-3}$, and $L=50\,\mu{\rm m}$. The point $P\,(6.3,\pi)$ (solid blue circle) in panel (a) is the one with $B=6.3$\,{\rm G} and $\phi_+=\pi$, while the point $Q\,(-6.3,\pi)$ in panel (b) is the one for $B=-6.3$\,{\rm G} and $\phi_-=\pi$.
From these results, we see that both $\eta_{j}$ and $\phi_{j}$ are very sensitive to $B$. Thereby, the present system can be used to design a magnetometer to detect the external magnetic field $B$, which can be realized by
measuring the amplitude attenuation factors $\eta_{j}$ and/or phase shifts $\phi_{j}$ of the photon polarization qubit.

\section{Discussion and summary}\label{Sec4}

Notice that the calculation results given above are based on the assumption that the gate atom locates at the fixed position $z=z_g$.
To be rigorous and realistic, the derivation of the above results by the influence of gate-atom delocalization (which may be due to the intrinsic quantum motion of the gate atom and also due to possible other noise acting on the atom) should be estimated. To this end, we assume that the gate atom may randomly occupy different spatial positions around $z_g$, with the density described by $\rho_g(z^{\prime}_g,\xi)= f\left(\xi\right) \delta[z^{\prime}_g-(z_g+\xi)]$.
Here $f(\xi)\equiv[1/(\sqrt{\pi} \sigma)] \exp [-(\xi / \sigma)^2]$ is normalized distribution function, with $\sigma$ the distribution width and $\xi$ the random variable describing the derivation of the gate-atom position relative to $z_g$.
Based on such a random density distribution, we have carried out a numerical  simulation on the topics described above, with the result presented in Appendix~\ref{app5}. It shows that the gate-atom delocalization gives no significant modification to the main conclusions given above, which means that the single-photon qubit switch, phase shifts, and weak magnetic-field measurement can still be achieved in the system.

In conclusion, in this article we have suggested and analyzed a scheme for manipulating the propagation of single photon pulses of two polarization components in a cold atomic gas via double Rydberg-EIT. Through solving the Heisenberg-Maxwell equations governing the quantum dynamics of the atoms and quantized probe field,
we have shown that, by storing a gate photon in a Rydberg state, a deep and adjustable optical potential for photon polarization qubits can be realized based on the strong Rydberg-Rydberg interaction. We have also shown that this scheme can be utilized to design all-optical switches of photon polarization qubits in the dissipative propagation regime, and generate large phase shifts to them in the dispersive propagation regime. Furthermore, we have demonstrated that such a scheme can be employed to detect weak magnetic fields that induce the Zeeman splitting of the atomic levels.

The theoretical approach developed here can be generalized to the study of all-optical transistors and phase gates of photon qubits and qudits based on Rydberg atoms. The results reported in this work are useful not only for the understanding of the quantum optical property of Rydberg atomic gases, but also for the design of novel quantum devices at single-photon levels, promising
in applications for quantum information processing.


\section*{Acknowledgments}
The authors thank Zhangyang Bai and Jingzhong Zhu for useful discussions.
This work was supported by the National Natural Science Foundation of China under Grant No.~11975098, and by the Research Funds of Happiness Flower ECNU under Grant No.~2020ECNU-XFZH005.

\appendix
\renewcommand{\thefigure}{S\arabic{figure}}
\setcounter{figure}{0}

\section{Explicit expression of the Heisenberg equation of motion (\ref{HL})}\label{app1}
Explicit expression of the Heisenberg equation of motion (\ref{HL}) for atomic operators is given by
\begin{widetext}
\begin{subequations}\label{HL_explicit}
\begin{align}
& i\left(\frac{\partial}{\partial t}+\Gamma_{21}\right)\hat{S}_{11}-i\Gamma_{12}\hat{S}_{22}-i\Gamma_{13}
\hat{S}_{33}+g_{p+}^\ast{\hat E}_{p+}^{\dag}\hat{S}_{31}-g_{p+}\hat{S}_{13}{\hat E}_{p+}-i{\hat F}_{11}=0,\\
& i\left(\frac{\partial}{\partial t}+\Gamma_{12}\right)\hat{S}_{22}-i\Gamma_{21}\hat{S}_{11}-i\Gamma_{23}
\hat{S}_{33}+g_{p-}^\ast{\hat E}_{p-}^{\dag}\hat{S}_{32}-g_{p-}\hat{S}_{23}{\hat E}_{p-}-i{\hat F}_{22}=0,\\
& i\left(\frac{\partial}{\partial t}+\Gamma_{3}\right)\hat{S}_{33}-i\Gamma_{34}\hat{S}_{44}-
g_{p+}^\ast{\hat E}_{p+}^{\dag}\hat{S}_{31}+g_{p+}{\hat E}_{p+}\hat{S}_{13}-g_{p-}^\ast{\hat E}_{p-}^{\dag}\hat{S}_{32}+g_{p-}\hat{S}_{23}{\hat E}_{p-}\nonumber\\
& +\Omega_{c}^\ast\hat{S}_{43}-\Omega_{c}\hat{S}_{34}-i{\hat F}_{33}=0,\\
& i\left(\frac{\partial}{\partial t}+\Gamma_{34}\right)\hat{S}_{44}-\Omega_{c}^\ast\hat{S}_{43}+\Omega_{c}\hat{S}_{34}-i{\hat F}_{44}=0,\\
& \left(i\frac{\partial}{\partial t}+d_{21}\right)\hat{S}_{21}+g_{p-}^{\ast}{\hat E}_{p-}^{\dag}\hat{S}_{31}-g_{p+}\hat{S}_{23}{\hat E}_{p+}-i{\hat F}_{21}=0, \\
& \left(i\frac{\partial}{\partial t}+d_{31}\right)\hat{S}_{31}+\Omega_{c}^\ast\hat{S}_{41}+g_{p+}(\hat{S}_{11}-\hat{S}_{33}){\hat E}_{p+}+g_{p-}\hat{S}_{21}{\hat E}_{p-}-i{\hat F}_{31}=0, \\
& \left(i\frac{\partial}{\partial t}+d_{32}\right)\hat{S}_{32}+\Omega_{c}^\ast\hat{S}_{42}
+g_{p-}(\hat{S}_{22}-\hat{S}_{33}){\hat E}_{p-}+g_{p+}\hat{S}_{12}{\hat E}_{p+}-i{\hat F}_{32}=0, \\
& \left(i\frac{\partial}{\partial t}+d_{41}-\Delta_{d}(z)\right)\hat{S}_{41}
+\Omega_{c}\hat{S}_{31}-g_{p+}\hat{S}_{43}{\hat E}_{p+}-i{\hat F}_{41}=0, \\
& \left(i\frac{\partial}{\partial t}+d_{42}-\Delta_{d}(z)\right)\hat{S}_{42}
+\Omega_{c}\hat{S}_{32}-g_{p-}\hat{S}_{43}{\hat E}_{p-}-i{\hat F}_{42}=0, \\
& \left(i\frac{\partial}{\partial t}+d_{43}-\Delta_{d}(z)\right)\hat{S}_{43}+\Omega_{c}\left(\hat{S}_{33}
-\hat{S}_{44}\right)-g_{p+}^\ast{\hat E}_{p+}^{\dag}\hat{S}_{41}-g_{p-}^\ast{\hat E}_{p-}^{\dag}\hat{S}_{42}-i{\hat F}_{43}=0.
\end{align}
\end{subequations}
\end{widetext}
Here
$d_{\alpha\beta}=\Delta_{\alpha}-\Delta_{\beta}+i\gamma_{\alpha\beta}$
($\alpha\neq \beta)$, $\gamma_{\alpha\beta}\equiv(\Gamma_\alpha+\Gamma_\beta)/2+\gamma_{\alpha\beta}^{\rm dep}$, and $\Gamma_\beta\equiv\sum_{\alpha<\beta}\Gamma_{\alpha\beta}$. $\Gamma_{\alpha\beta}$ is the decay rate of the spontaneous emission from the state $|\beta\rangle$ to the state $|\alpha\rangle$, $\gamma_{\alpha\beta}^{\rm dep}$ is the dephasing rate between $|\alpha\rangle$ and $|\beta\rangle$.
The half Rabi frequency of the control field is defined by $\Omega_c\equiv(\mathbf{e}_c\cdot \mathbf{p}_{43})\mathcal{E}_c/\hbar$.

\section{Derivation of the two-component envelope equations of the probe field}\label{app2}
The dynamical evolution of the probe field is controlled by the HM equations (\ref{HL}) and (\ref{Maxwell2}). Because we are interested in the case of the probe field at a single-photon level, the nonlinear terms in the HM equations play no significant role and hence can be disregarded safely. Based on this idea, we
take ${\hat S}_{\alpha\beta}\rightarrow {S}_{\alpha\beta}^{(0)}+ {\hat S}_{\alpha\beta}$,
with ${S}_{\alpha\beta}^{(0)}$ the steady-state solution of ${\hat S}_{\alpha\beta}$ in the absence of the probe field (i.e.   ${S}_{11}^{(0)}={S}_{22}^{(0)}=1/2$ and other ${S}_{\alpha\beta}^{(0)}=0$). Then by taking ${\hat S}_{\alpha\beta}$ and
$\hat{E}_{pj}$ as small quantities, Eqs.~(\ref{HL}) and (\ref{Maxwell2}) are reduced into
\begin{subequations}\label{HL_linearized}
\begin{align}
\label{H1}
& \left(i\frac{\partial}{\partial t}+d_{31}\right){\hat S}_{31}+\Omega_{c}^{\ast}{\hat S}_{41}+\frac{g_{p+}{\hat E}_{p+}}{2}-i{\hat F}_{31}=0,\\
\label{H2}
& \left(i\frac{\partial}{\partial t}+d_{32}\right){\hat S}_{32}+\Omega_{c}^{\ast}{\hat S}_{42}+\frac{g_{p-}{\hat E}_{p-}}{2}-i{\hat F}_{32}=0,\\
\label{H3}
& \left[i\frac{\partial}{\partial t}+d_{41}-\Delta_{d}(z)\right]{\hat S}_{41}+\Omega_{c}{\hat S}_{31}-i{\hat F}_{41}=0,\\
\label{H4}
& \left[i\frac{\partial}{\partial t}+d_{42}-\Delta_{d}(z)\right]{\hat S}_{42}+\Omega_{c}{\hat S}_{32}-i{\hat F}_{42}=0, \\
\label{M1}
& i\left(\frac{\partial}{\partial z}+\frac{1}{c}\frac{\partial}{\partial t}\right){\hat E}_{p+}
+\frac{g_{p+}^{\ast}N}{c}{\hat S}_{31}=0,\\
\label{M2}
& i\left(\frac{\partial}{\partial z}+\frac{1}{c}\frac{\partial}{\partial t}\right){\hat E}_{p-}
+\frac{g_{p-}^{\ast}N}{c}{\hat S}_{32}=0.
\end{align}
\end{subequations}
Since these equations are linear, they can be solved easily by using the Fourier transform:
\begin{subequations}\label{Fourier transform}
\begin{align}
\hat{X}(z, t) &=\frac{1}{\sqrt{2 \pi}} \int_{-\infty}^{+\infty} d \omega \tilde{\hat{X}}(z, \omega) e^{-i \omega t}, \\
\tilde{\hat{X}}(z, \omega)&=\frac{1}{\sqrt{2 \pi}} \int_{-\infty}^{+\infty} d t \hat{X}(z, t) e^{i \omega t},
\end{align}
\end{subequations}
where $\hat{X}$ denotes ${\hat S}_{31}$, ${\hat S}_{32}$, ${\hat S}_{41}$, ${\hat S}_{42}$, ${\hat F}_{31}$, ${\hat F}_{32}$, ${\hat F}_{41}$, ${\hat F}_{42}$, and ${\hat E}_{pj}$ ($j=+, -$). Substituting (\ref{Fourier transform}) into  (\ref{H1})-(\ref{H4}), we get the atomic transition operators expressed by the polarization components of the probe field:
\begin{subequations}
\begin{align}
\label{S31}
\tilde{\hat{S}}_{31}&=\frac{Y_{1}(\omega)}{2D_{1}(\omega)}g_{p+}\tilde{\hat E}_{p+}
-i\frac{Y_{1}(\omega) \tilde{\hat{F}}_{31}-\Omega_c \tilde{\hat{F}}_{41}}{D_{1}(\omega)} ,\\
\label{S32}
\tilde{\hat{S}}_{32}&=\frac{Y_{2}(\omega)}{2D_{2}(\omega)}g_{p-}\tilde{\hat E}_{p-}
-i\frac{Y_{2}(\omega) \tilde{\hat{F}}_{32}-\Omega_c \tilde{\hat{F}}_{42}}{D_{2}(\omega)} ,\\
\label{S41}
\tilde{\hat{S}}_{41}&=\frac{-\Omega_{c}}{2D_{1}(\omega)}g_{p+}\tilde{\hat E}_{p+}
+i\frac{\Omega_{c}\tilde{\hat{F}}_{31}
-(\omega+d_{31})\tilde{\hat{F}}_{41}}{D_{1}(\omega)}, \\
\label{S42}
\tilde{\hat{S}}_{42}&=\frac{-\Omega_{c}}{2D_{2}(\omega)}g_{p-}\tilde{\hat E}_{p-}
+i\frac{\Omega_{c}\tilde{\hat{F}}_{32}-(\omega+d_{32})
\tilde{\hat{F}}_{42}}{D_{2}(\omega)},
\end{align}
\end{subequations}
where $D_{\alpha}(\omega)=|\Omega_{c}|^2-(\omega+d_{3\alpha})[\omega+d_{4\alpha}
-\Delta_{d}(z)]$, $Y_{\alpha}(\omega)=\omega+d_{4\alpha}-\Delta_{d}(z)$  ($\alpha = 1,2$).

\begin{figure*}
\includegraphics[width=2.1\columnwidth]{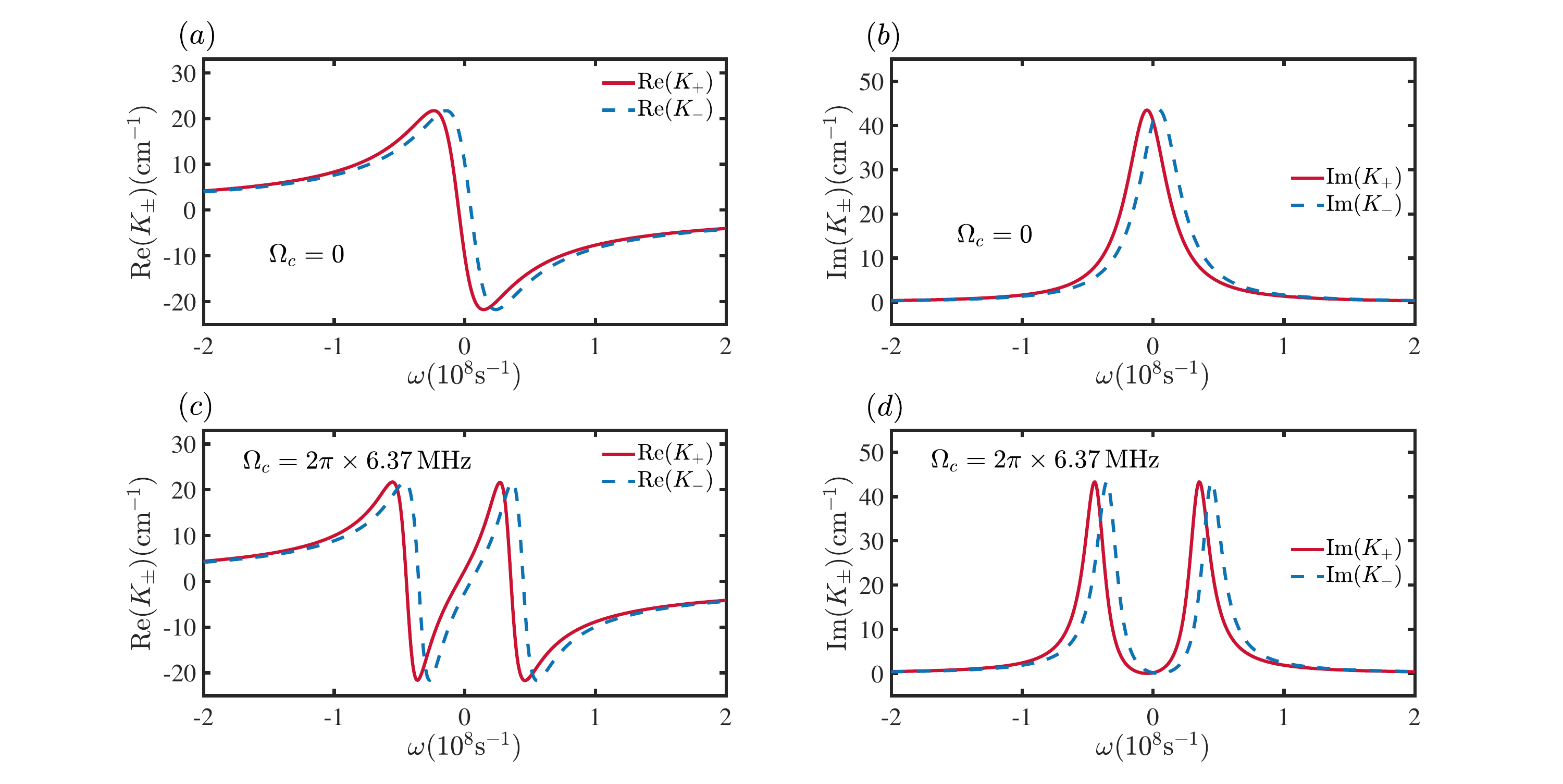}  
\caption{{\footnotesize
Linear dispersion relations $K_{+}$ (for $\sigma^{+}$ component) and $K_{-}$ (for $\sigma^{-}$ component) of the probe field as functions of $\omega$ in the absence of the gate atom, by taking $\Delta_{2}=-\Delta_{1}=4.68$\,MHz ($B=1.6$\,{\rm G}), $\Delta_{3}=\Delta_{4}=0$, and ${\cal N}_{a}=3 \times 10^{10}\,{\rm cm}^{-3} $.
(a),(b)~Real parts ${\rm Re}(K_{\pm})$ and imaginary parts ${\rm Im}(K_{\pm})$  for $\Omega_{c}=0$. No EIT occurs in this case.
(c),(d) The same as panels (a) and (b), but with  $\Omega_c=2\pi\times6.37$\,MHz.
An EIT transparency window is opened in both ${\rm Im}(K_{+})$ and ${\rm Im}(K_{-})$, i.e. a double EIT occurs.
}}
\label{FigS1}
\end{figure*}

Substituting  (\ref{S31}) and (\ref{S32})
into (\ref{M1}) and (\ref{M2}), we obtain the two-component envelope equations:
\begin{align}\label{LSE}
\left[i\frac{\partial}{\partial z}+K_{j}(z,\omega)\right]{\tilde{\hat E}}_{pj}(z,\omega)=i{\tilde{\hat{{\cal F}}}}_{pj}(z,\omega),
\end{align}
$j=+,-$. Here
\begin{subequations}
\begin{align}
&& K_+(z,\omega)=\frac{\omega}{c}+\frac{|g_{p}|^{2}N}{2c}
\frac{[\omega+d_{41}-\Delta_{d}(z)]}{D_{1}(\omega)},\\
&& K_-(z,\omega)=\frac{\omega}{c}+\frac{|g_{p}|^{2}N}{2c}
\frac{[\omega+d_{42}-\Delta_{d}(z)]}{D_{2}(\omega)},
\end{align}
\end{subequations}
are linear dispersion relations of the $\sigma^{+}$  and $\sigma^{-}$ component of the probe field, respectively;
${\tilde{\hat{{\cal F}}}}_{pj}(z,\omega)$ are defined by
\begin{subequations}\label{LBF}
\begin{align}
\tilde{\hat{\mathcal F}}_{p +}(z, \omega)=\frac{g_{p +}^* N}{c} \frac{Y_{1}(\omega) \tilde{\hat F}_{31}(z, \omega)-\Omega_c \tilde{\hat F}_{41}(z, \omega)}{D_{1}(\omega)}, \label{LBF1}\\
\tilde{\hat{\mathcal F}}_{p -}(z, \omega)=\frac{g_{p -}^* N}{c} \frac{Y_{2}(\omega) \tilde{\hat F}_{32}(z,\omega)-\Omega_c \tilde{\hat F}_{42}(z, \omega)}{D_{2}(\omega)}. \label{LBF2}
\end{align}
\end{subequations}

Note that in the above derivation, for simplicity, the quantity $\Delta_d (z)$  has been assumed to be a slowly-varying function of $z$, which allows us to take it as a constant during the Fourier transformation~\cite{noteapp2}. The quantity $\omega$ in the Fourier transformation (\ref{Fourier transform}) plays a role of the sideband angular frequency of the probe pulse (the center angular frequency is $\omega_p$).
Under the EIT condition, the Langevin noise terms ${\tilde{\hat{{\cal F}}}}_{pj}(z,\omega)$ are very small and hence can be neglected safely (see detailed discussions about the role of Langevin noise in EIT systems given in Refs.~\cite{Gorshkov2011PRL,Zhu2021,Zhu2022,Zhu2023}).

\section{Double Rydberg-EIT for $\Delta_d(z)=0$}\label{app3}

If the gate atom is absent, the position-dependent detuning $\Delta_d(z)=0$.
In this case, from (\ref{FT2}) and (\ref{FT3}) we have $K_{+}(z,\omega)\rightarrow K_{+}(\omega)$, $K_{-}(z,\omega)\rightarrow K_{-}(\omega)$, with
\begin{subequations}\label{0FT}
\begin{align}
K_{+}(\omega)=\frac{\omega}{c}+\frac{|g_{p}|^{2}N}{2c}
\frac{\omega+d_{41}}{|\Omega_{c}|^2-(\omega+d_{31})(\omega+d_{41})},\label{0FT1}\\
K_{-}(\omega)=\frac{\omega}{c}+\frac{|g_{p}|^{2}N}{2c}
\frac{\omega+d_{42}}{|\Omega_{c}|^2-(\omega+d_{32})(\omega+d_{42})}. \label{0FT2}
\end{align}
\end{subequations}

Fig.~\ref{FigS1}
shows $K_{+}$ (for $\sigma^{+}$ component) and $K_{-}$ (for $\sigma^{-}$ component) as functions of $\omega$, plotted by taking $\Delta_{2}=-\Delta_{1}=4.68$\,MHz ($B=1.6$\,{\rm G}), $\Delta_{3}=\Delta_{4}=0$, and ${\cal N}_{a}=3 \times 10^{10}\,{\rm cm}^{-3} $.
From the figure, we see that no EIT occurs for the both polarization components if the control field is absent (i.e., $\Omega_{c}=0$); see the single-peak absorption spectra ${\rm Im}(K_{+})$ and ${\rm Im}(K_{-})$ shown in Fig.~\ref{FigS1}(b). However, when the control field is applied ($\Omega_c=2\pi\times6.37$\,MHz), an EIT transparency window is opened in both
${\rm Im}(K_{+})$ and ${\rm Im}(K_{-})$; see the two-peak absorption spectra plotted in Fig.~\ref{FigS1}(d). This means that a double EIT occurs in the present inverted-Y system. Especially, when $B=0$, the two polarization components are nearly degenerate (and hence the level configuration is symmetric), $K_{+}$ and $K_{-}$ are nearly coincided with each other.

\section{Derivation of the Rydberg defect potential}\label{app4}

The preparation of the gate atom results in a position-dependent detuning $\Delta_d(z)$, which induces a Rydberg defect potential for the propagation of the probe pulse. To show this, we note that Eq.~(\ref{RLSE}) in the main text can be written as the form (when neglecting the Langevin noise terms)
\begin{eqnarray}\label{RLSEA}
&& i\hbar\frac{\partial}{\partial \tau} \tilde{\hat E}_{pj}(z,\omega)=V_{j}(z,\omega)\tilde{\hat E}_{pj}(z,\omega),
\end{eqnarray}
with $\tau\equiv ct$ and $V_{j}(z,\omega)\equiv -\hbar cK_{j}(z,\omega)$ ($j=+,-$). One sees that $V_{+}(z,\omega)$ and $V_{-}(z,\omega)$ act as roles of external potentials for the $\sigma^+$ and $\sigma^-$ polarization components, respectively. Obviously, the $z$-dependence of $V_{\pm}(z,\omega)$ is a reflection of the Rydberg defect potential.

Here, for simplicity, we give a detailed discussion on $V_{\pm}(z,\omega)$ near the center point of the EIT transparency windows (i.e. at $\omega=0$). Based on the result of (\ref{FT2}) and (\ref{FT3}), we have $V_{\pm}(z,0)\equiv V_{\pm}(z) = -\hbar c K_{\pm}(z, 0) = {\rm Re}[V_{\pm}(z)]+i\,{\rm Im}[V_{\pm}(z)]$, with the detailed expressions given by

\begin{widetext}
\begin{subequations}\label{VA}
\begin{align}
& {\rm Re}[V_{+}(z)]=\frac{{\cal N}_{a}\omega_{p}|{\bf e}_{p+}\cdot{\bf p}_{31}|^2}{4\varepsilon_{0}}\frac{[\Delta_{d}(z)+\Delta_{1}][|\Omega_{c}|^2
+(\Delta_{3}-\Delta_{1})[\Delta_{d}(z)+\Delta_{1}]]+(\Delta_{3}-\Delta_{1})
\gamma_{41}^{2}-[\Delta_{d}(z)+\Delta_{1}]\gamma_{31}\gamma_{41}}
{\left||\Omega_{c}|^2-d_{31}[d_{41}-\Delta_{d}(z)]\right|^2},\label{VA1}\\
& {\rm Re}[V_{-}(z)]=\frac{{\cal N}_{a}\omega_{p}|{\bf e}_{p-}\cdot{\bf p}_{32}|^2}{4\varepsilon_{0}}\frac{[\Delta_{d}(z)+\Delta_{2}][|\Omega_{c}|^2
+(\Delta_{3}-\Delta_{2})[\Delta_{d}(z)+\Delta_{2}]]+(\Delta_{3}-\Delta_{2})
\gamma_{42}^{2}-[\Delta_{d}(z)+\Delta_{2}]\gamma_{32}\gamma_{42}}
{\left||\Omega_{c}|^2-d_{32}[d_{42}-\Delta_{d}(z)]\right|^2},\label{VA2}\\
& {\rm Im}[V_{+}(z)]=-\frac{{\cal N}_{a}\omega_{p}|{\bf e}_{p+}\cdot{\bf p}_{31}|^2}{4\varepsilon_{0}}\frac{[\Delta_{d}(z)+\Delta_{1}]^{2}\gamma_{31}
+|\Omega_{c}|^2\gamma_{41}+\gamma_{31}\gamma_{41}^{2}}{\left||\Omega_{c}|^2
-d_{31}[d_{41}-\Delta_{d}(z)]\right|^2},\label{VA3}\\
& {\rm Im}[V_{-}(z)]=-\frac{{\cal N}_{a}\omega_{p}|{\bf e}_{p-}\cdot{\bf p}_{32}|^2}{4\varepsilon_{0}}\frac{[\Delta_{d}(z)+\Delta_{2}]^{2}\gamma_{32}
+|\Omega_{c}|^2\gamma_{42}+\gamma_{32}\gamma_{42}^{2}}{\left||\Omega_{c}|^2
-d_{32}[d_{42}-\Delta_{d}(z)]\right|^2}.\label{VA4}
\end{align}
\end{subequations}
\end{widetext}
In deriving the above formula, we have set the two-photon detuning $\Delta_{4}=0$, which is required for obtaining significant EIT effect.

\section{Influence due to the gate-atom delocalization}\label{app5}

The calculation results presented in Sec.~\ref{Sec3} of the main text are obtained based on the assumption that the gate atom locates exactly at a fixed position $z=z_g=L/2$. This is, strictly speaking, hard to achieve since one cannot determine the exact position of the gate atom due to the intrinsic quantum motion of the atom and also due to possible other noise acting on the atom~\cite{Murray2016,Thompson2017,Pohl2019}. To be rigorous and also realistic, the influence of gate-atom delocalization should be considered.

To estimate the deviation due to the gate-atom delocalization, we assume that the gate atom may occupy different spatial positions around $z_g$ in random way. This can be described by the random density distribution of the gate atom with the form $\rho_g(z^{\prime}_g,\xi)= f\left(\xi\right) \delta[z^{\prime}_g-(z_g+\xi)]$.
Here $f(\xi)\equiv[1/(\sqrt{\pi} \sigma)] \exp [-(\xi / \sigma)^2]$ is a normalized statistical distribution function ($\int_{-\infty}^{\infty}f(\xi)=1$), with $\sigma$ being the distribution width and $\xi$ being a random variable describing the gate-atom coordinate deviated from the center position $z=z_g$.
Hence the position-dependent detuning $\Delta_{d}(z)$ given in (\ref{Delta0}) is changed into the form $\Delta_{d}(z, \xi)=-f\left(\xi\right)\frac{C_6}{\left|(z_g+\xi)-z\right|^{6}}$.

\begin{figure*}
\centering
\centerline{\includegraphics[width=2.3\columnwidth]{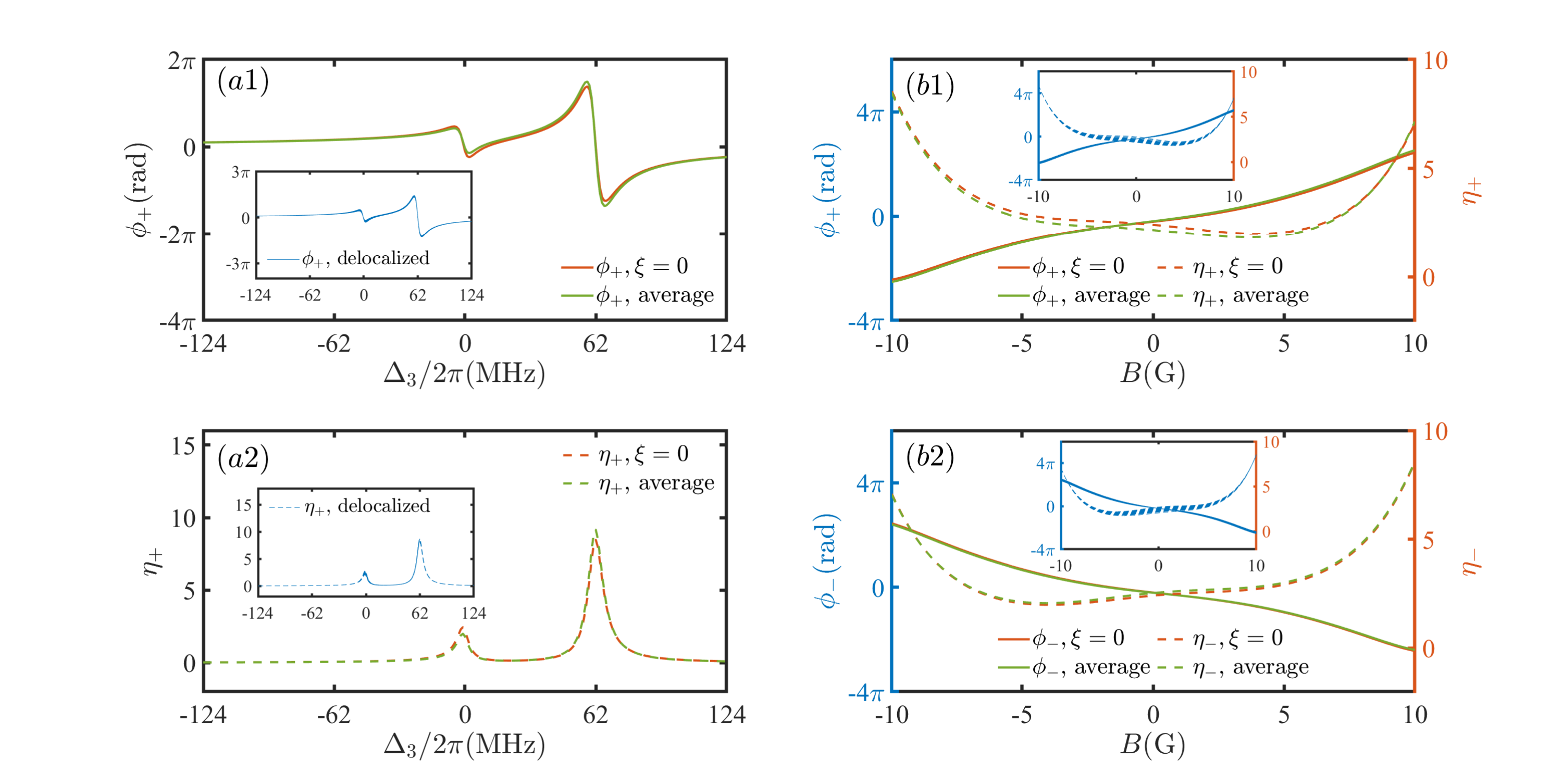}}
\caption{{\footnotesize
The influence of the gate-atom delocalization.
(a1)~Phase shift $\phi_+$ of the $\sigma^+$ component of the qubit wavepacket as a function of one-photon detuning $\Delta_3$.
Green curve is the result of the statistical average by taking 200 different random $\xi$ values ($\xi$ is a random variable describing the coordinate deviation of the gate atom from the center position $z=z_g=L/2$); red line is the one for $\xi=0$, corresponding to the gate atom fixed at $z=z_g$.
Blue curves in the inset are results by taking different values of $\xi$.
(a2)~Similar to (a1), but for the amplitude attenuation factor of $\eta_{+}$ of the qubit wavepacket.
(b1)~Phase shift $\phi_+$ and amplitude attenuation $\eta_{+}$ of the $\sigma^+$ component as functions of the magnetic field $B$.
(b2)~Similar to (b1) but for $\phi_-$  and $\eta_{-}$ of the $\sigma^-$ component.
For (a1,a2), system parameters used are $L=80\,\mu$m, $B=1.4$\,{\rm G}, ${\cal N}_{a}=3\times10^{12}\,{\rm cm}^{-3}$; while for (b1,b2), they are given by $\Delta_3=0$, ${\cal N}_{a}=3\times10^{12}\,{\rm cm}^{-3}$, and $L=100\,\mu{\rm m}$.
One sees that the gate-atom delocalization has no significant influence on
the qubit switch, phase shifts, and magnetic-field measurement in the system.
}}
\label{FigS2}
\end{figure*}

Fig.~\ref{FigS2} shows the result of the numerical simulation
on the influence of the gate-atom delocalization to the qubit switch, phase shift, and magnetic-field measurement in the system. Illustrated in panel
(a1) is the phase shift $\phi_+$ of the $\sigma^+$ polarization component of the qubit wavepacket as a function of one-photon detuning $\Delta_3$. The
green curve in the figure is the statistical average of $\phi_+$ by taking 200 different values of $\xi$, while the red line is the one for $\xi=0$, which corresponds to the case where the gate atom is fixed at $z=z_g$.
Blue curves in the inset of the figure is the result (describing the fluctuations of $\phi_+$) by taking different values of $\xi$. System parameters used are $L=80\,\mu$m, $B=1.4$\,{\rm G}, ${\cal N}_{a}=3\times10^{12}\,{\rm cm}^{-3}$. Panel (a2) gives the result for the amplitude attenuation factor $\eta_{+}$ (the parameter describing the switch behavior of the system) of the qubit wavepacket.
In the simulation, 200 different $\xi$ values are chosen between $30$\,$\mu$m and $50$\,$\mu$m. Since $\phi_-$ and $\eta_{-}$ of the $\sigma^-$ component have similar behaviors as the $\sigma^+$ component, they are not shown here.

Illustrated in panel (b1) is the result of the phase shift $\phi_+$ and amplitude attenuation factor $\eta_{+}$  as functions of magnetic field $B$.
Red and green lines in the figure is for $\xi=0$ and the one for the statistical average by taking 200 random $\xi$ values, respectively. The inset of the figure (where blue curves denoting fluctuations are plotted)  is the result obtained by taking different values of $\xi$. System parameters chosen here are $\Delta_3=0$, ${\cal N}_{a}=3\times10^{12}\,{\rm cm}^{-3}$, $L=100\,\mu{\rm m}$, with  $40\,\mu{\rm m} \le \xi\le 60\,\mu{\rm m}$.
Shown in panel (b2) is similar to that of panel (b1), but for the phase shift $\phi_-$  and the amplitude attenuation factor  $\eta_{-}$ of the $\sigma^-$ polarization component.

By inspecting the four panels (a1), (a2), (b1), and (b2), we
see that the green curves (the results of the statistical average for many different random values of gate-atom position) are very closed to the red ones (the result for the fixed position of the gate atom),
which means that the influence caused by gate-atom delocalization is small and has no qualitative impact on the main conclusions given in the main text. Thereby, the single-photon qubit switch, phase shifts, and weak magnetic-field measurement can still be achieved in the system even in the presence of gate-atom delocalization.


\begin{thebibliography}{99}
\newcommand{\enquote}[1]{``#1''}

\bibitem{Boyd2008}
R. W. Boyd,
\textit{Nonlinear Optics} (3rd edition)
(Academic, New York, 2008).

\bibitem{Fleischhauer2005}
M. Fleischhauer, A. Imamoglu, and J. P. Marangos,
Electromagnetically Induced Transparency: Optics in Coherent Media,
Rev. Mod. Phys. {\bf 77}, 633 (2005).

\bibitem{Chang2014}
D. E. Chang, V. Vuleti\'{c}, and M. D. Lukin,
Quantum nonlinear optics - photon by photon,
Nat. Phys. {\bf 8}, 685 (2014), and references therein.


\bibitem{Gallagher2008}
T. F. Gallagher,
{\it Rydberg Atoms} (Cambridge Univ. Press, Cambridge, England, 2008).

\bibitem{Saffman2010}
M. Saffman, T. G. Walker, and K. M$\phi$lmer,
Quantum information with Rydberg atoms,
Rev. Mod. Phys. {\bf 82}, 2313 (2010), and references therein.


\bibitem{Friedler2005}
 I. Friedler, D. Petrosyan, M. Fleischhauer, and G. Kurizki,
 Long-range interactions and entanglement of slow single-photon pulses,
 Phys. Rev. A. {\bf 72}, 043803 (2005).

\bibitem{Mohapatra2007PRL}
A. K. Mohapatra, T. R. Jackson, and C. S. Adams,
Coherent Optical Detection of Highly Excited Rydberg States Using Electromagnetically Induced Transparency, Phys. Rev. Lett. {\bf 98}, 113003 (2007).

\bibitem{Adams2010}
J. D. Pritchard, D. Maxwell, A. Gauguet, K. J. Weatherill, M. P. A. Jones and C. S. Adams,
Cooperative atom-light interaction in a blockaded Rydberg ensemble,
Phys. Rev. Lett. {\bf 105}, 193603 (2010).



\bibitem{Gorshkov2011PRL}
A. V. Gorshkov, J. Otterbach, M. Fleischhauer, T. Pohl, and M. D. Lukin,
Photon-Photon Interactions via Rydberg blockade,
Phys. Rev. Lett. {\bf 107}, 133602 (2011).

\bibitem{Firstenberg2013}
O. Firstenberg, T. Peyronel, Q.-Y. Liang, A. V. Gorshkov, M. D. Lukin, V. Vuleti\'c,
Attractive photons in a quantum nonlinear medium,
Nature {\bf 502}, 71 (2013).

\bibitem{He2014}
B. He, A. V. Sharypov, J. Sheng, C. Simon, and M. Xiao,
Two-Photon Dynamics in Coherent Rydberg Atomic Ensemble,
Phys. Rev. Lett. {\bf 112}, 133606 (2014).

\bibitem{Bienias2014}
P. Bienias, S. Choi, O. Firstenberg, M. F. Maghrebi, M. Gullans, M. D. Lukin,  A. V. Gorshkov, and H. P. B\"uchler,
Scattering resonances and bound states for strongly interacting Rydberg polaritons,
Phys. Rev. A {\bf 90}, 053804 (2014).

\bibitem{Caneva2015}
T. Caneva, M. T Manzoni, T. Shi, J. S. Douglas1, J. I. Cirac, and
D. E. Chang,
Quantum dynamics of propagating photons with strong interactions:
a generalized input–output formalism,
New J. Phys. {\bf 17}, 113001 (2015).

\bibitem{Maghrebi2015}
M. F. Maghrebi, M. J. Gullans, P. Bienias, S. Choi, I. Martin, O. Firstenberg,
M. D. Lukin, H. P. B\"uchler, and A. V. Gorshkov,
Coulomb Bound States of Strongly Interacting Photons,
Phys. Rev. Lett. {\bf 115}, 123601 (2015).

\bibitem{Li2015}
W. Li and I. Lesanovsky,
Coherence in a cold-atom photon switch,
Phys. Rev. A {\bf 92}, 043828 (2015).

\bibitem{Gullans2016}
M. J. Gullans, J. D. Thompson, Y. Wang, Q.-Y. Liang, V. Vuleti\'c, M. D. Lukin, and A. V. Gorshkov,
Effective Field Theory for Rydberg Polaritons,
Phys. Rev. Lett. {\bf 117}, 113601 (2016).


\bibitem{Murray2016}
C. R. Murray, A. V. Gorshkov, and T. Pohl,
Many-body decoherence dynamics and optimized operation of a single-photon switch,
New J. Phys. {\bf  18}, 092001 (2016).


\bibitem{Thompson2017}
J. D. Thompson, T. L. Nicholson, Q. Liang, S. H. Cantu, A. V. Venkatramani, S. Choi, I. A. Fedorov, D. Viscor, T. Pohl, M. D. Lukin and V. Vuleti\'c,
Symmetry-protected collisions between strongly interacting photons,
Nature {\bf 542}, 206 (2017).

\bibitem{Pohl2019}
M. Khazali, C. R. Murray and T. Pohl,
Polariton Exchange Interactions in Multichannel Optical Network,
Phys. Rev. Lett. {\bf 123}, 113605 (2019).


\bibitem{Jachymski2016}
K. Jachymski, P. Bienias, and H. P. B\"uchler,
Three-Body Interaction of Rydberg Slow-Light Polaritons,
Phys. Rev. Lett. {\bf 117}, 053601 (2016).

\bibitem{Das2016}
S. Das, A. Grankin, I. Iakoupov, E. Brion, J. Borregaard, R. Boddeda, I. Usmani, A. Ourjoumtsev, P. Grangier, and A. S. S\o rensen,
Photonic Controlled-Phase Gates through Rydberg Blockade in Optical Cavities,
Phys. Rev. A. {\bf 93}, 040303(R)(2016).

\bibitem{Yang2016}
L. Yang, B. He, J.-H. Wu, Z. Zhang, and M. Xiao,
Interacting photon pulses in a Rydberg medium,
Optica {\bf 3}, 1095 (2016).

\bibitem{Gullans2017}
M. J. Gullans, S. Diehl, S. T. Rittenhouse, B. P. Ruzic, J. P. D'Incao, P. Julienne, A. V. Gorshkov, and J. M. Taylor,
Efimov States of Strongly Interacting Photons,
Phys. Rev. Lett. {\bf 119}, 233601 (2017).

\bibitem{Moos2017}
M. Moos, R. Unanyan, and M. Fleischhauer,
Creation and detection of photonic molecules in Rydberg gases,
Phys. Rev. A {\bf 96}, 023853 (2017).

\bibitem{Liang2018}
Q.-Y. Liang, A. V. Venkatramani, S. H. Cantu, T. L. Nicholson,
M. J. Gullans, A. V. Gorshkov, J. D. Thompson, C. Chin,
M. D. Lukin, V. Vuleti\'c,
Observation of three-photon bound states in a quantum nonlinear medium,
Science {\bf 359}, 783 (2018).

\bibitem{Cantu2020}
S. H. Cantu, A. V. Venkatramani, W. Xu, L. Zhou, B. Jelenkovi\'c,
M. D. Lukin, and V. Vuleti\'c,
Repulsive photons in a quantum nonlinear medium,
Nat. Phys. {\bf 16}, 921 (2020).

\bibitem{Bienias2020}
P. Bienias, M. J. Gullans, M. Kalinowski,  A. N. Craddock,
D. P. Ornelas-Huerta, S. L. Rolston, J. V. Porto, and A. V. Gorshkov,
Exotic Photonic Molecules via Lennard-Jones-like Potentials,
Phys. Rev. Lett. {\bf 125}, 093601 (2020).

\bibitem{Ou2022}
Y. Ou, Q. Zhang, and G. Huang,
Quantum reflection of single photons in a cold Rydberg atomic gas,
Opt. Lett. {\bf 47}, 4395 (2022).

\bibitem{Drori2023}
L. Drori, B. C. Das, T. D. Zohar, G. Winer, E. Poem, A. Poddubny, and O. Firstenberg,
Quantum vortices of strongly interacting photons,
Science {\bf 381}, 193 (2023).

\bibitem{DingY2023}
Y. Ding, Z. Bai, G. Huang, and W. Li,
Facilitation-Induced Transparency and Single-Photon Switch with Dual-Channel
Rydberg Interactions,
Phys. Rev. Applied  {\bf 19}, 014017 (2023).

\bibitem{Murray2016adv}
C. Murray and T. Pohl,
Quantum and nonlinear optics in strongly interacting atomic ensembles,
in {\it Advances in Atomic, Molecular, and Optical Physics}
(Academic, New York, 2016), Vol. 65, pp. 321-372,
and references therein.

\bibitem{Adams2020}
C. S. Adams, J. D. Pritchard, and J. P. Shaffer,
Rydberg atom quantum technologies,
J. Phys. B: At. Mol. Opt. Phys. {\bf 53}, 012002 (2020),
and references therein.


\bibitem{Dudin2012}
Y. O. Dudin and A. Kuzmich,
Strongly Interacting Rydberg Excitations of a Cold Atomic Gas,
Science {\bf 336}, 887 (2012).

\bibitem{Peyronel2012}
T. Peyronel, O. Firstenberg, Q.-Y. Liang, S. Hofferberth, A. V. Gorshkov, T. Pohl,
M. D. Lukin, V. Vuleti\'c,
Quantum nonlinear optics with single photons enabled by strongly interacting atoms,
Nature {\bf 488}, 57 (2012).

\bibitem{Baur2014PRL}
S. Baur, D. Tiarks, G. Rempe, and S. D\"{u}rr,
Single-Photon Switch Based on Rydberg Blockade,
Phys. Rev. Lett. {\bf 112}, 073901 (2014).

\bibitem{Gorniaczyk2014}
H. Gorniaczyk, C. Tresp, J. Schmidt, H. Fedder, and S. Hofferberth,
Single-Photon Transistor Mediated by Interstate Rydberg Interactions,
Phys. Rev. Lett. {\bf 113}, 053601 (2014).

\bibitem{Tiarks2014PRL}
D. Tiarks, S. Baur, K. Schneider, S. D\"{u}rr, and G. Rempe,
Single-Photon Transistor Using a F\"orster Resonance,
Phys. Rev. Lett. {\bf 113}, 053602 (2014).

\bibitem{Gorniaczyk2016}
H. Gorniaczyk, C. Tresp, P. Bienias, A. Paris-Mandoki, W. Li, I. Mirgorodskiy, H.P. B\"uchler, I. Lesanovsky, and S. Hofferberth,
Enhancement of Rydberg-mediated single-photon nonlinearities by electrically tuned F\"orster resonances,
Nat. Commun. {\bf 7}, 12480 (2016).

\bibitem{Tiarks2016}
D. Tiarks, S. Schmidt, G. Rempe, and S. D\"{u}rr,
Optical $\pi$ phase shift created with a single-photon pulse,
Sci. Adv. {\bf 2}, 1600036 (2016).

\bibitem{Ripka2018}
F. Ripka, H. K\"ubler, R. L\"ow, T. Pfau,
A room-temperature single-photon source based on strongly interacting
Rydberg atoms,
Science {\bf 362}, 446 (2018).

\bibitem{Tiarks2019}
D. Tiarks, S. Schmidt-Eberle, T. Stolz, G. Rempe, and S. D\"{u}rr,
A photon-photon quantum gate based on Rydberg interactions,
Nat. Phys. {\bf 15}, 124 (2019).

\bibitem{Ornelas-Huerta2020}
D. P. Ornelas-Huerta, A. N. Craddock, E. A. Goldschmidt, A. J. Hachtel,
Y. Wang, P. Bienias, A. V. Gorshkov, S. L. Rolston, and J. V. Porto,
On-demand indistinguishable single photons from an efficient and pure source based on a Rydberg ensemble,
Optica {\bf 7}, 813 (2020).

\bibitem{Vaneecloo2022}
J. Vaneecloo, S. Garcia, and A. Ourjoumtsev,
Intracavity Rydberg Superatom for Optical Quantum Engineering: Coherent Control, Single-Shot Detection, and Optical $\pi$ Phase Shift,
Phys. Rev. X. {\bf 12} 021034 (2022)

\bibitem{Stolz2022}
T. Stolz, H. Hegels, M. Winter, B. R\:{o}hr, Y.-F. Hsiao, L. Husel,
G, Rempe, and S, D\"{u}rr,
Quantum-Logic Gate between Two Optical Photons with an Average Efficiency above $40\%$,
Phys. Rev. X {\bf 12}, 021035 (2022).

\bibitem{Shi2022}
S. Shi, B. Xu, K. Zhang, G.-S. Ye, D.-S. Xiang, Y. Liu, J. Wang, D. Su, and  L. Li,
High-fidelity photonic quantum logic gate based on near-optimal Rydberg single-photon source,
Nat. Commun. {\bf 13}, 4454 (2022).

\bibitem{Ye2023}
G.-S. Ye, B. Xu, Y. Chang, S. Shi, T. Shi, and  L. Li,
A photonic entanglement filter with Rydberg atoms,
Nat. Photon. {\bf 17}, 538 (2023).


\bibitem{Sinclair2019}
J. Sinclair, D. Angulo, N. Lupu-Gladstein, K. Bonsma-Fisher, and A. M. Steinberg, Observation of a Large, Resonant, Cross-Kerr Nonlinearity in a Free-space Rydberg Medium, Phys. Rev. Research {\bf 1}, 0331931 (2019).


\bibitem{Mu2021}
Y. Mu, L. Qin, Z. Shi, and G. Huang,
Giant Kerr nonlinearities and magneto-optical rotations in a Rydberg-atom gas via double electromagnetically induced transparency,
Phys. Rev. A {\bf 103}, 043709 (2021).

\bibitem{Shi2021PRA}
Z. Shi and G. Huang,
Self-organized structures of two-component laser fields and their active
control in a cold Rydberg atomic gas,
Phys. Rev. A {\bf 104}, 013511 (2021).

\bibitem{Shi2021OL}
Z. Shi and G. Huang,
Selection and cloning of periodic optical patterns with a cold Rydberg atomic gas,
Opt. Lett. {\bf 47}, 6221 (2021).

\bibitem{Mu2022}
Y. Mu and G. Huang,
Stern–Gerlach effect of vector light bullets in a nonlocal Rydberg medium,
Opt. Lett. {\bf 46}, 5344 (2022).


\bibitem{Kok2010}
P. Kok and W. B. Lovett, {\it Introduction to Optical Quantum Information Processing} (Cambridge Univ. Press, Cambridge, England, 2010).


\bibitem{Petrosyan2004}
D. Petrosyan and Y. P. Malakyan,
Magneto-optical rotation and cross-phase modulation via coherently driven four-level atoms in a tripod configuration,
Phys. Rev. A. {\bf 70}, 023822 (2004).


\bibitem{note00}
The two polarization components has no spatial separation during propagation. The separation plotted in the figure is used to indicate the fact that the photon qubit consists of two polarization components.


\bibitem{note0}
The radius of the Rydberg blockade sphere $r_b$ $\equiv (|C_6| |d_{31}|/|\Omega_c|^2)^{1/6}$ is about $8\, \mu {\rm m}$. To make the (1+1)-dimensional model be valid,  the condition $\pi r_b^2>A_0$ must be satified.


\bibitem{noteapp2}
Using the system parameters, one can obtain the group velocity of the probe pulse $V_g\sim 10^{-7}c$. If the time duration of the pulse is $t_0\sim 10^{-7}$\,s, the spatial width of the pulse along the $z$-direction is given by $d_0\sim V_g t_0\sim 3\,\mu$m, which is much less the spatial width of $\Delta_d (z)$ ($\sim 10\,\mu$m). This means that during the Fourier transformation $\Delta_d (z)$ can indeed be approximated as a constant.


\bibitem{Zhu2021}
J. Zhu, Q. Zhang, and G. Huang,
Quantum squeezing of slow-light solitons,
Phys. Rev. A {\bf 103}, 063512 (2021).

\bibitem{Zhu2022}
J. Zhu and G. Huang,
``Quantum squeezing of slow-light dark solitons via electromagnetically induced transparency,''
Phys. Rev. A {\bf 105}, 033515 (2022).

\bibitem{Zhu2023}
J. Zhu, Y. Mu, and G. Huang,
Simultaneous quantum squeezing of light polarizations and atomic spins in a cold atomic gas,
Phys. Rev. A {\bf 107}, 033517 (2023).


\bibitem{Miller2010}
D. A. B. Miller,
Are optical transistors the logical next step?
Nat. Photon. {\bf 4}, 3 (2010).

\bibitem{note1}
Since $\gamma_{41}$ and $\gamma_{42}$ are much smaller than $\Delta_d(z)$, the contribution to the amplitude attenuations and phase shifts by $d_{41}$ and $d_{42}$ in (\ref{it11}) and (\ref{it12}) are negligible.


\bibitem{Nielsen2000}
M. A. Nielsen and I. L. Chuang,
{\it Quantum Computation and Quantum Information} (Cambridge
Univ. Press, Cambridge, 2000).

\bibitem{Budker2013}
D. Budker and D. F. K. Rochester (eds.), Optical Magnetometry (Cambridge, London, 2013).






%
%
%
%
%


\end{thebibliography}
\end{document}